\documentclass[12pt, a4paper, twoside]{article}
\usepackage{a4wide}
\usepackage[dvips]{graphics}
\usepackage{epsfig}
\usepackage{eurosym}

\def\geenr	{{$^{76}$Ge}}
\def\cosix	{{$^{60}$Co}}
\def\cum	{{m$^3$}}
\def\radzzs     {{$^{226}$Ra}}
\def\thzza      {{$^{228}$Th}}
\def\kvn        {{$^{40}$K}}
\def\gam        {{$\gamma$}}
\def\ln         {LN}
\def\lar        {LAr}
\def\ctsper     {cts/(keV$\cdot$kg$\cdot$y)}
\def\dctsper     {{$10^{-3}$~cts/(keV$\cdot$kg$\cdot$y)}}
\def\vctsper     {{$10^{-4}$~cts/(keV$\cdot$kg$\cdot$y)}}

\def\etal       {{\it et al.}}
\def\NIM        {Nucl.~Instr.~Meth.}
\def\PRB        {Phys.~Rev.~B}
\def\borex      {{\mbox{{\sc Borexino}}}}
\def\geni      {{\mbox{{\sc Genius}}}}
\def\CERN      {{\mbox{{\sc Cern}}}}
\def\GEANT      {{\mbox{{\sc Geant}}}}
\def\LVD      {{\mbox{{\sc Lvd}}}}
\def\GNO      {{\mbox{{\sc Gno}}}}
\def\IGEX      {{\mbox{{\sc Igex}}}}
\def\LNGS      {{\mbox{{\sc Lngs}}}}
\def\CUORI      {{\mbox{{\sc Cuoricino}}}}
\def\CUORE      {{\mbox{{\sc Cuore}}}}
\def\Majorana      {{\mbox{{\sc Majorana}}}}
\def\AGATA      {{\mbox{{\sc Agata}}}}
\def\WMAP      {{\mbox{{\sc Wmap}}}}
\def\NEMO      {{\mbox{{\sc Nemo}}}}
\def\KATRIN      {{\mbox{{\sc Katrin}}}}
\def\MOON      {{\mbox{{\sc Moon}}}}
\def\EXO      {{\mbox{{\sc Exo}}}}
\def\GEM      {{\mbox{{\sc Gem}}}}
\def\CAMEO      {{\mbox{{\sc Cameo}}}}
\def\COBRA      {{\mbox{{\sc Cobra}}}}

\def\balf {\rule{10mm}{2mm}}

\newfont{\ord}{cmsy10 scaled 1200}

\begin{document}

%
%
%
%
 
\begin{titlepage}

\def\LNGS     {\,{\bf $^{a}$, }} 
\def\JINR     {\,{\bf $^b$, }} 
\def\HD       {\,{\bf $^c$, }}
\def\KO       {\,{\bf $^d$, }}
\def\MIL      {\,{\bf $^e$, }} 
\def\INR      {\,{\bf $^f$, }} 
\def\ITEP     {\,{\bf $^g$, }} 
\def\KU       {\,{\bf $^h$, }} 
\def\MU       {\,{\bf $^i$, }} 
\def\TU       {\,{\bf $^j$, }} 
\def\JINRINR  {\,{\bf $^{b,f}$, }}
\def\LNGSMIL  {\,{\bf $^{a,e}$, }} 

\def\lngs     {{$^a$}\,} 
\def\jinr     {{$^b$}\,} 
\def\hd       {{$^c$}\,}
\def\ko       {{$^d$}\,}
\def\mil      {{$^e$}\,}  
\def\inr      {{$^f$}\,} 
\def\itep     {{$^g$}\,} 
\def\ku       {{$^h$}\,} 
\def\mu       {{$^i$}\,} 
\def\tu       {{$^j$}\,}


\begin{center}

{\LARGE\bf A New $^{76}$Ge Double Beta Decay Experiment}
\vskip0.5truecm
{\LARGE\bf                  at LNGS}

\vskip4truecm

\includegraphics[width=14cm]{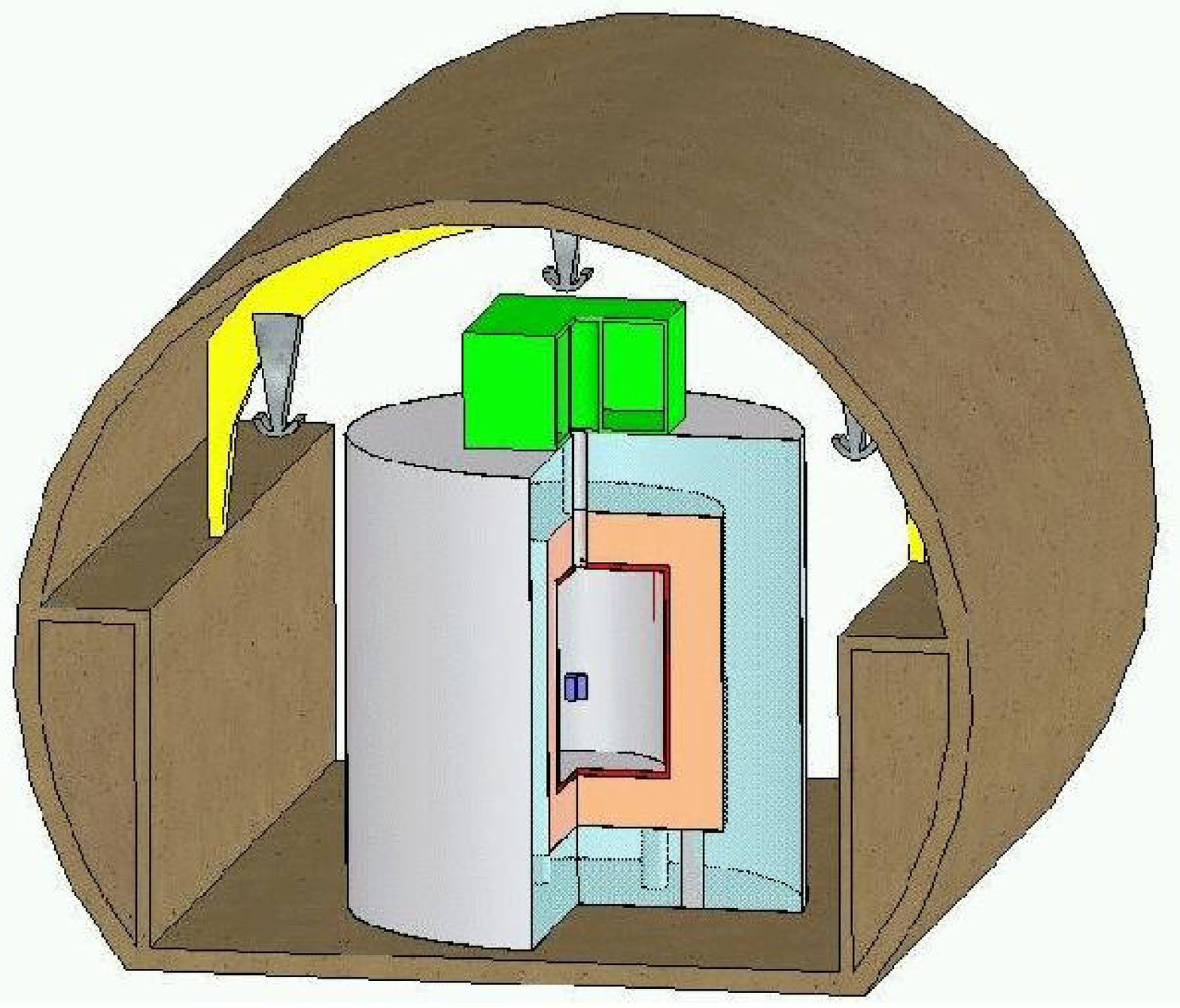}

\vskip2truecm

\vfill

{\LARGE\bf Letter of Intent}
\vskip0.5truecm

{\bf 16 March 2004}

\end{center}
\newpage

\begin{center}
{\LARGE\bf A New $^{76}$Ge Double Beta Decay Experiment}
\vskip0.5truecm
{\LARGE\bf                  at LNGS}
\end{center}

\vskip15truemm
\noindent
\begin{center}
I.~Abt\MU
M.~Altmann\MU 
A.~Bakalyarov\KU
I.~Barabanov\INR
C.~Bauer\HD
E.~Bellotti\MIL
S.T.~Belyaev\KU
L.~Bezrukov\INR
V.~Brudanin\JINR
C.~B\"uttner\MU
V.P.~Bolotsky\ITEP 
A.~Caldwell\MU
C.~Cattadori\LNGSMIL
H.~Clement\TU
A.~Di Vacri\LNGS
J.~Eberth\KO
V.~Egorov\JINR
G.~Grigoriev\KU
V.~Gurentsov\INR 
K.~Gusev\JINR
W.~Hampel\HD
G.~Heusser\HD
W.~Hofmann\HD
J.~Jochum\TU
M.~Junker\LNGS
J.~Kiko\HD
I.V.~Kirpichnikov\ITEP
A.~Klimenko\JINRINR
K.T.~Kn\"opfle\HD
V.N.~Kornoukhov\ITEP
M.~Laubenstein\LNGS
V.~Lebedev\KU
X.~Liu\MU
I.~Nemchenok\JINR
L.~Pandola\LNGS
V.~Sandukovsky\JINR
S.~Sch\"onert\HD
S.~Scholl\TU
B.~Schwingenheuer\HD
H.~Simgen\HD
A.~Smolnikov\JINRINR
A.~Tikhomirov\KU
A.A.~Vasenko\ITEP
S.~Vasiliev\JINRINR  
D.~Wei\ss haar\KO
E.~Yanovich\INR
J.~Yurkowski\JINR 
S.~Zhukov\KU
G.~Zuzel$^d$      	
\end{center}

\vskip1truecm

\begin{center}

\def\vhere {\vskip2truemm}

{\rm \lngs INFN Laboratori Nazionali del Gran Sasso, Assergi, Italy} \\
\vhere
{\rm \jinr Joint Institute for Nuclear Research, Dubna, Russia}  \\
\vhere
{\rm \hd Max-Planck-Institut f\"ur Kernphysik, Heidelberg, Germany} \\
\vhere
{\rm \ko Institut f\"ur Kernphysik, Universit\"at K\"oln, Germany}  \\
\vhere
{\rm \mil Universit\`a di Milano Bicocca e INFN Milano, Milano, Italy}  \\
\vhere
{\rm \inr Institute for Nuclear Research of the Russian Academy of Sciences, Moscow, Russia} \\
\vhere
{\rm \itep Institute for Theoretical and Experimental Physics, Moscow, Russia}\\
\vhere
{\rm \ku Russian Research Centre Kurchatov Institute, Moscow, Russia}                         \\
\vhere
{\rm \mu Max-Planck-Institut f\"ur Physik, M\"unchen, Germany}         \\
\vhere
{\rm \tu Physikalisches Institut, Universit\"at T\"ubingen, Germany}  \\

\vfill

{\underbar{Contact Persons}}:\\ \vskip1truemm
 W.~Hofmann \\
{\em (\,Werner.Hofmann@mpi-hd.mpg.de\,) }\\ \vskip1truemm
 S.~Sch\"onert, \\
{\em (\,Stefan.Schoenert@mpi-hd.mpg.de\,) }\\  
\end{center}

\vskip0.5truecm

\end{titlepage}

\tableofcontents
\newpage
\section{Overview and executive summary}
{\bf The physics case.}
Since their discovery neutrinos have been an object of extensive
experimental study and the knowledge about their properties has
advanced our understanding of weak interactions significantly.
Still unanswered, however, is the very fundamental question whether
the neutrino is a Majorana particle like most extensions of the
Standard Model assume. 
The study of neutrinoless double beta decay is the most
sensitive approach to answer this question and the potential of this
method has increased considerably during the last years since a
non-zero mass of the neutrinos has been established by the  
observation of neutrino flavor oscillation. In fact, the
observation of neutrinoless double beta decay would not only establish the
Majorana nature of the neutrino but also provide a measurement
of its effective mass. 

Recently, the group of Klapdor-Kleingrothaus
has claimed for the first time positive indications
for neutrinoless double beta decay of $^{76}$Ge \cite{kla04}. In this
Letter of Intent we propose a new facility at the  Laboratori Nazionali del Gran Sasso
(\LNGS )  
 which will allow to study decays of $^{76}$Ge at
unprecedented low background levels.\\

\noindent
{\bf The experimental situation.} At present the most sensitive
experiments -- Heidelberg-Moscow and \IGEX\ --
use $^{76}$Ge as source and detector, and reach sensitivities
around 0.3~eV in the effective neutrino mass. Both collaborations have
reported almost the same
upper limit on the lifetime of 1.6$\cdot$10$^{25}$~y, corresponding 
to a mass range of 0.33 to 1.3~eV. However, the 
group of Klapdor-Kleingrothaus claims a 4~$\sigma$ excess in the spectrum near
the energy expected for neutrinoless double beta decay, and gives 
a neutrino mass range from 0.2~eV to 0.6~eV. 
Currently,
no other experiment can either confirm
or refute this observation. Closest to this reach are cosmological limits based on
the amplitude of higher frequencies in the spatial distribution of galaxies;
streaming of mass-less neutrinos should reduce the spectral power at high frequency.
\WMAP\ has given upper mass limits of 0.23~eV per species \cite{spe03}. Other authors have
pointed out that this value is model-dependent,
and that more conservative estimates give higher values \cite{elg03}.
However, data have also been interpreted as showing positive evidence of massive
neutrinos \cite{all03}. 
Double beta decay experiments now taking data such as \CUORI\ \cite{arn03}
or \NEMO\,\cite{nemo}
may reach the 0.3~eV sensitivity region within a few years, using other
nuclei as sources. These experiments have the potential to confirm the
current positive evidence with similar significance, but cannot refute it
because of uncertainties in the ratio of nuclear matrix elements. On a 
similar time scale the \KATRIN\ experiment \cite{osi01} will reach 
a mass sensitivity of 0.2~eV to 0.3~eV  
based on the study of the tritium beta decay spectrum near the
end point. A positive signature by \KATRIN\ together with a sufficiently
sensitive experiment on neutrinoless double beta decay would establish the
Dirac or Majorana nature of neutrinos. Regardless of the outcome of these
experiments, it is clear that a $^{76}$Ge experiment capable of confirming
the current result with high significance, or pushing the mass limits below
the current, cosmologically still important 0.3~eV range is of high
scientific relevance.\\ \clearpage

\noindent
{\bf The new low-level facility based on a cryogenic fluid shield.}
In an experiment searching for neutrinoless double beta decay, the number
of decays expected is proportional to the detector mass $M$ and the 
measurement time $T$. The crucial factor determining the performance of 
the experiment is the background in the mass window of the neutrinoless
double beta decay line. As long as there are no background counts, lifetime
limits improve as 
$(MT)$ whereas in the presence of background 
limits go as $(MT)^{1/2}$. 
Thus, background suppression is the key to a successful experiment. Background
spectra in current $^{76}$Ge experiments point to the shielding material
as the dominant source of residual background. Therefore, it was proposed
already several years ago to operate bare germanium diodes in a shield
of liquid nitrogen \cite{heu95}; the \geni\ and \GEM\ proposals \cite{kla99,zde01} were based on this idea. 
Since nitrogen can be purified extremely well from
radioactive contaminants, it is one of the radio-purest environments
possible and should allow a background reduction by two to three orders
of magnitude compared to past $^{76}$Ge experiments.
This  allows 
background-free measurements up to lifetimes well beyond 10$^{26}$ years.
Reliable operation of germanium detectors in liquid nitrogen and liquid argon
has been demonstrated in several laboratory experiments and, on a larger
scale, in the \geni\ Test Facility \cite{kla03a}.
Additional suppression of external and internal backgrounds
can be achieved by vetoing techniques, using veto signals from other
nearby detectors, from other segments of the same detector in case of
segmented detectors, from the pulse shape which indicates multiple interaction
sites within a detector or detector segment, or from scintillation light
in the surrounding liquid in case liquid argon is employed
instead of liquid nitrogen.

We propose to install a novel facility based on an
(optionally active)
cryogenic fluid shield in Hall A of the Gran Sasso Laboratory. The 
facility serves a dual purpose. 
The setup allows to
scrutinize with high significance on a short time scale the current evidence for neutrinoless
double beta decay of $^{76}$Ge using the existing $^{76}$Ge detectors
from the previous Heidelberg-Moscow and \IGEX\ experiments.
An increase in the lifetime limit can be achieved by adding more enriched
detectors, remaining thereby background-free up to a few 100 kg-years of 
exposure.
On the other hand, it is a pioneering
low-level facility in that it allows to
develop and test low-background measurements with Germanium
detectors at background levels several orders of magnitude below
the current state-of-the-art; it represents a major step on the way
towards ultimate double beta decay experiments aiming for a sensitivity
in the 10~meV mass range. 

For cost and space reasons, we consider to use a combination of shields,
rather than a single thick liquid nitrogen shield as in the original
\geni\ proposal \cite{kla99}. While the optimum shield configuration is still under
investigation, a promising layout is to use a 1.5~m liquid nitrogen shield
followed by a $\approx 10$~cm shield of high-purity lead still inside the
cryostat. The lead shields activity from the cryostat walls and insulation
material and results in a compact cryostat, important both for cost and
safety aspects, since the volume of the cryogenic fluid is kept relatively small. The 
thickness of the liquid is chosen to sufficiently shield remaining activities in the
lead. An outer shield of roughly 2~m of water complements the shielding
against the rock and concrete. It also serves as a neutron shield and --
encased in a diffuse reflecting foil and equipped with photomultipliers --
as a veto against cosmic muons. A cleanroom and sophisticated lock and 
suspension systems on top of the cryostat allow to insert and remove detectors 
without introducing contamination into the vessel. Gas purification and
handling systems make extensive use of the experience accumulated in
\borex . Signal recording and pulse shape analysis make use of recent
developments where signals are sampled and digitized at high 
rate after minimal analog signal processing, providing maximal information
and flexibility for a later digital post-processing.\\

\noindent
{\bf Phases of the experiment.}
The experiment would proceed in several phases. \\ \noindent
Phase I encompasses the installation of the cryostat and shields,
the installation and operation of conventional Ge detectors to
determine the background rejection and to screen materials 
and identify backgrounds by
classifying their spectra, and the operation of almost 20~kg of existing enriched
$^{76}$Ge detectors, used in the past in the Heidelberg-Moscow and \IGEX\
experiments. 
Within one year of measurement,
the sensitivity of this setup should allow a statistically
unambiguous statement concerning neutrinoless double beta decay
with a lifetime around 1.2$\cdot$10$^{25}$~y, as measured by \cite{kla04}.

Phase II: In parallel with the data taking of the first phase, techniques will
by studied and implemented to provide improved enriched detectors to
be used in a second phase. Enriched germanium will be produced in Russia;
two different options to enrich the germanium are under discussion.
Detector geometry and segmentation will be optimized on the basis of
detailed calculations of fields and pulse shapes, 
taking into account background
simulations. Particular emphasis is devoted to minimize cosmogenic
activation of detectors by reducing the exposure and optionally moving
production steps to underground facilities. Regardless of the
outcome of the Phase I measurements, it is desirable to produce and
operate a certain number of new detectors: 
In case of a positive result to provide a precise lifetime measurement, 
in case of a negative outcome to push the limits further. 
In particular in the second case, one would -- funding permitting --
add enriched detectors up to the point where backgrounds start
to show up.  

At the end of Phase II with 100 kg$\cdot$years, the sensitivity will be
T$_{1/2} > 2\cdot 10^{26}$ years at 90\% confidence level (C.L.) corresponding to a range of the
effective neutrino mass of $<$0.09\,-\,0.29~eV.

{Phase III:} The ultimate experiment capable to reach the 10~meV scale requires
{\ord O}(1~t) of enriched germanium and represents another huge step, which
can only be afforded in the context of a world-wide collaboration.
Options for detector shielding and detector arrangements will have to be
reevaluated on the basis of results achieved by the proposed experiment
and by studies following other approaches, such as the copper shield
foreseen in the Majorana proposal \cite{maj03}.
It is likely that by that time the proposed facility will have 
reached its limits - in fiducial size, in background shielding, or in both --
and needs to be upgraded or replaced by an improved facility.
On the scale of the cost of 1~t of enriched germanium, costs for such
a new facility are modest. At the current time, it is, however, clearly
premature to speculate about the kind of modifications needed, or the
space required by and the potential location of such a future experiment.
\\ \clearpage

\noindent
{\bf Time scale, cost, and requests to \LNGS .}
Making use of the available know-how concerning gas purification in
\borex , and ideally also of the \borex\ infrastructure, we estimate
that the cryostat and auxiliary system could be set up on a time
scale of less than 2 years. After installation
of the $^{76}$Ge detectors a measurement time of at least three 
years is required. The cost of the cryostat, the auxiliary systems and
the modification of the existing detectors is estimated to be 3~M\euro .

Phase II -- the production of new enriched detectors -- could start
concurrently with Phase I. Less than 3 years will be required
until the first detectors are available. The cost of Phase II depends
on the amount of additional detectors, and on the production mode.
Detector costs are roughly 100~\euro /g, including the raw material,
the enrichment and the crystal growing, plus a certain offset since
the crystal growing requires a certain amount of additional material
(which can be recovered for a second lot of detectors). Underground
detector preparation would require laboratory equipment worth about
1~M\euro .

The facility for Phase I and II could be located in the free space
in Hall A of \LNGS . Electronics, experiment control and gas control
could be housed partly on top of the cryostat, partly in (stacked) containers. 
Additional space is required for liquid gas tanks. 
Safety issues regarding the cryogenic fluid
system and gas handling need to be addressed in close collaboration
with \LNGS\ safety officers.\\

\newpage

\section{Introduction and experimental overview}
        
\subsection{Introduction}

Double beta decays are transitions between nuclei of the same atomic mass number ($A$)
that change the nuclear charge
$(Z)$ by two units
under emission of light particles. Double beta decay is only observable
in  absence of the concurring process, the cascading decay via two 
single beta decays. 
This condition is only satisfied, if the mass of the intermediate 
nucleus is larger than that of the initial one, or if the 
single beta transition to the intermediate nucleus is highly hindered.
Double beta transitions for both signs of nuclear charge change 
are possible: two neutrons transform into two 
protons, or vice versa, two protons into two neutrons. 
For simplicity, we consider here
only the first.

The transformation can occur under emission of two neutrinos
($\beta\beta(2\nu)$),  
\begin{equation}
(A,Z) \rightarrow (A,Z + 2) + e^-_1 + e^-_2 + {\bar\nu}_{e1} + 
{\bar\nu}_{e2},         
\end{equation}
conserving lepton number.
In contrast, the neutrinoless decay $(\beta\beta(0\nu))$
\begin{equation}
(A,Z) \rightarrow (A,Z + 2) + e^-_1 + e^-_2     
\end{equation}
violates lepton number by two units and is forbidden in the 
standard electroweak theory. A further decay mode involves 
the emission of a light neutral boson ($\beta\beta(0\nu,\chi)$), 
a majoron, as postulated 
in some extensions of the standard electroweak theory: 
\begin{equation}
(A,Z) \rightarrow (A,Z + 2) + e^-_1 + e^-_2 + \chi 
\end{equation}

The different decay modes are distinguishable by the shape of 
the spectrum of the electron sum energy. For the 
$\beta\beta(2\nu)$ mode, the summed kinetic energy of the 
two electrons displays a continuous spectrum with a broad maximum
below half the endpoint energy. 
In contrast,  the $\beta\beta(0\nu)$ mode  
exhibits a mono-energetic line at the endpoint ($Q_{\beta\beta}$), 
as the electrons
carry the full available energy. For a light majoron,
the spectrum is also continuous with a broad peak located
above half the endpoint energy.

Neutrinoless double beta decay can be mediated by various mechanism. 
Here we consider only the simplest case of the 
left-handed $V-A$ weak currents and the exchange of a 
light massive Majorana neutrino.
The half life is then (e.g. \cite{ell02})
\begin{equation}
[T^{0\nu}_{1/2}(0^+ \rightarrow 0^+)]^{-1} =
G^{0\nu}(E_0,Z) |M^{0\nu}_{GT} - g^2_V / g^2_A M^{0\nu}_F|^2  m_\nu^2~,
\end{equation} 
where $G^{0\nu}$ is the phase-space integral, $M^{0\nu}_{GT}, M^{0\nu}_F$ are
the nuclear matrix elements, and $m_\nu$ the effective electron neutrino mass.
Under the assumption of three light 
massive Majorana neutrinos $\nu_i \,\, (i=1,2,3)$, the weak eigenstate neutrinos $\nu_e,
\nu_\mu$ and $\nu_\tau$ can be written as a superposition of the mass
eigenstates $\nu_i$ with the mixing matrix
$U_{li}$. The electron neutrino $\nu_e $ 
is then given as $\nu_e = \sum_{i}^{3} U_{ei} \nu_i$ and
the effective neutrino mass defined as
\begin{equation}
 m_\nu^2 = \left| \sum_{i}^{3}U^2_{ei}m_i \right| ^2 = 
\left| \sum_{i}^{3}|U_{ei}|^2 e^{\alpha_i}m_i \right| ^2,
\end{equation}
including  two CP violating Majorana phases $\alpha_i$ 
which can cause cancellations in the sum. 

\begin{figure}[ht]
\begin{center}
\includegraphics[width=12cm]{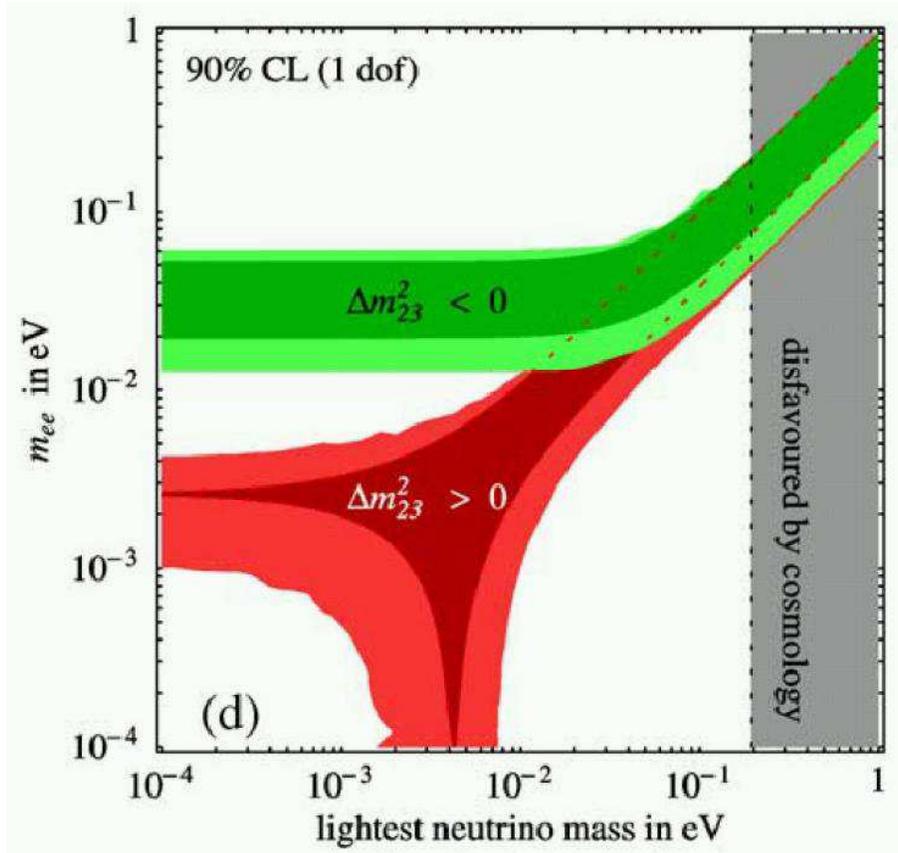}
\end{center}
\caption{\label{fig:me_vs_m1}
Predictions for the effective neutrino mass 
$m_\nu$ as a function of the lightest neutrino mass $m_1$ derived 
from oscillation experiments \cite{fer03}. The different bands in parameter
space correspond to the normal mass hierarchy ($\Delta m_{23} > 0$), 
to the inverted 
mass hierarchy ($\Delta m_{23} < 0$)  and to the degenerate mass scheme
(lightest mass $\gg \Delta m_{23}$).
The dark colors
use the measured oscillation data without errors to emphasize
the contribution of the Majorana CP phases. The lighter colors include
the current experimental errors of the mixing parameters.
}
\end{figure}

From the measurements of the mass differences 
$\Delta m^2_{ij} =|m_i^2 - m_j^2|$ and the mixing angles in neutrino 
oscillation experiments, the range for $m_\nu$ is substantially constrained.
Figure~\ref{fig:me_vs_m1} displays the range of $m_\nu$ as a function 
of the lightest neutrino mass $m_1$.
One distinguishes the
parameter range corresponding to  normal, inverted 
and degenerated mass schemes. 
In the normal hierarchy, the mass splitting driving solar neutrino 
oscillations occurs between the lightest neutrino $m_1$ and   $m_2$,
and that driving atmospheric oscillations between $m_2$ and   $m_3$.
In the inverted hierarchy the splitting is arranged
in the reverse order.
If the values of   $\Delta m^2_{ij}$ are small compared to
the actual values of $m_i$, the mass spectrum is called degenerate.
Large parts of the parameter space for $m_\nu^2$  predicted 
by neutrino oscillation experiments will be experimentally 
accessible with the next generation of experiments.

\subsection{Past and present experiments}

Major experimental progress has been achieved during the last ten years.
For a comprehensive review the reader is referred to \cite{ell02}. 
Direct measurements of double beta decay accompanied by the emission of 
two neutrinos ($\beta\beta(2\nu)$) 
have been carried out for more than ten nuclei. 
Recent results for $\beta\beta(2\nu)$ include 
$^{76}$Ge, $^{100}$Mo, $^{150}$Nd, 
$^{116}$Cd and $^{96}$Zr.
The measured half-lives are in the range of $10^{19} - 10^{21}$ years. 
 
Today the central focus in double beta decay research is the 
neutrinoless mode ($\beta\beta(0\nu)$). Most stringent
limits have been derived from experiments using 
enriched \geenr\ detectors.
The leading two experiments - \IGEX\ and Heidelberg-Moscow (HdM) - 
have been running for several years with a background 
around $Q_{\beta\beta}$ close to 0.2 \ctsper\ 
before pulse shape discrimination and  about 0.06 \ctsper\ after. 
Limits for the life time close to $2\cdot 10^{25}$~y have been 
derived corresponding to a limit for an effective neutrino mass $m_\nu$ of 
0.3~-~1.0~eV \cite{aal99,kla01}. Both experiments have stopped data 
taking recently.

In 2001, the group of Klapdor-Kleingrothaus claimed  evidence for
neutrinoless double beta decay at a 2~$\sigma$ confidence level  
\cite{kla01}
based on 52 (kg$\cdot$y) of data from the HdM experiment. 
Based on a data set of 72 (kg$\cdot$y) the claim has been strengthened recently 
\cite{kla04}. The derived 
access counts are $28.8\pm 6.9$ events above a background
of approximately 60 events.

Operational experiments are currently  \NEMO 3 and \CUORI .
The \NEMO\ experiment
is carried out at the Frejus 
Underground Laboratory, France. During five years of R\&D phase using
the \NEMO 2 detector, the collaboration has performed 
measurements of the $\beta\beta(2\nu)$ decay of $^{100}$Mo, 
$^{82}$Se, $^{116}$Cd and $^{96}$Zr. The final detector, \NEMO 3 \cite{nemo}, consists of a 
large tracking calorimeter surrounding 10 kg of thin source foils 
of different enriched materials, mainly 7 kg of $^{100}$Mo. 
The aim is 
to reach after five years of data taking 
a limit for the half-life of $5\cdot 10^{24}$ years, 
corresponding to an effective  mass  of 0.2-0.3 eV. 
\NEMO 3 started data taking in 2003. 

The \CUORI\ experiment at \LNGS\ searches for neutrinoless 
double beta decay with TeO$_2$ bolometers \cite{arn03}. The setup consists 
of an array of 62 crystals with a total mass of about 40 kg.
The counting rate in the region of neutrinoless double beta decay is 
$\sim$0.2~\ctsper . No evidence for neutrinoless 
double beta decay is found with the present exposure of about
three months during 2003. The corresponding limit for the lifetime
is $5.5\cdot 10^{23}$~y (90\% C.L.) and for the effective neutrino
mass between 0.37 and 1.9~eV \cite{arn04}. The expected 
sensitivity after three years of data taking will be $4\cdot 10^{24}$~y
or 0.2-0.5~eV \cite{giu03}.

\subsection{Proposed and suggested future experiments}

The next generation of double beta decay experiments  aims for probing 
Majorana masses down to 0.1 eV and below. Many different isotopes and 
detector concepts have been suggested. Recent reviews of the 
field can be found in \cite{ell02,giu03}. Here, we briefly discuss the 
more advanced projects as listed in Tab.~\ref{tab:futexp} including
this proposed new $^{76}$Ge experiment at \LNGS .

\begin{table}[htb]
\centering
\caption{{Characteristics of operating and proposed future $\beta\beta(0\nu)$ experiments.} 
The corresponding references are: \NEMO 3 \cite{aug03}, \CUORI\,\cite{giu03}, 
\NEMO -Next \cite{aug03}, 
\CUORE\,\cite{arn03},
\Majorana\ \cite{maj03}, \EXO\,\cite{dan00}.
The three phases of the proposed $^{76}$Ge 
experiment are discussed in the 
following sections. 
}
\label{tab:futexp}
\begin{tabular}{lrlcccc}
\noalign{\bigskip}
\hline
\hline
{Experiment} & {Source} & {Description} & {FWHM} & \multicolumn{2}{c}{Sensitivity}    & {Year}   \\
           &        &             & at $Q_{\beta\beta}$ & $T^{0\nu}_{1/2}$  
& $m_\nu$ & \\
 & & &(keV)&(y) & (eV) \\    
\hline
\hline
\NEMO 3     & $^{100}$Mo~ & 7 kg $^{enr}$Mo tracking  & 90  & $5\cdot 10^{24}$ 
& 0.2-0.3 & 2008 \\
\CUORI\  & $^{130}$Te~ & 40 kg TeO$_2$ bolom. & 7 & $4\cdot 10^{24}$ 
& 0.2-0.5 & 2007\\
\hline
\NEMO -Next  &  $^{100}$Mo~ & 0.1 t $^{enr}$Mo track.  & 50  & $1\cdot 10^{26}$
& 0.04-0.07 \\ 
\CUORE\  & $^{130}$Te~  & 0.76 t TeO$_2$ bolom. & 5  & $3\cdot 10^{26}$ 
& 0.03-0.05 \\
\Majorana\ & $^{76}$Ge~ & 0.5 t $^{enr}$Ge diodes & 4  & $4\cdot 10^{27}$
& 0.02-0.07 &  \\ 
\EXO\ & $^{136}$Xe~ & 1 t $^{enr}$Xe & 120  &  $8\cdot 10^{26}$ & 0.05-0.14 
\\
\hline
This Project & $^{76}$Ge~ & $^{enr}$Ge in LN/LAr & 4 & & & \\
~~~~Phase I      &           & 15 kg (15 kg y)    & & $3\cdot 10^{25}$  & 0.3-0.9 & 2006\\
~~~~Phase II      &           & 35 kg (100 kg y)  & & $2\cdot 10^{26}$  & 0.09-0.29    & 2009 \\
~~~~Phase III  & &  \multicolumn{3}{l}{{\ord O}(500~kg)~~~ -~~~ world-wide collaboration}  \\
\noalign{\smallskip}
\hline
\hline
\end{tabular}
\end{table}

\NEMO -Next would be based on the \NEMO 3 tracking 
concept, however with an increased mass of approximately 
100 kg of foils enriched in $^{100}$Mo and improved
energy resolution. A sensitivity of 
$>10^{26}$~y or of $0.04 - 0.07$~eV for  $m_\nu$ is projected.

The Cryogenic Underground Observatory for Rare Events (\CUORE ) 
has been proposed to be operated at the Gran Sasso laboratory.
It is planned to use 1000 crystals of TeO$_2$ with a total mass of 760 kg
as cryogenic bolometers \cite{arn03}. 
The detector is arranged into 25 separate towers
of 40 crystals. A prototype tower is operated 
in \CUORI . 
Assuming an energy resolution of 5~keV and a background of 
0.01(0.001) \ctsper , 
the expected sensitivity of \CUORE\ 
is $0.9(3.0)\cdot 10^{26} \sqrt{t}$ years.
One year of measurements would provide bounds
for $m_\nu $ in the $0.04 - 0.15$~eV range \cite{arn03}.

The \Majorana\ experiment plans to employ 500 kg of Ge, isotopically
enriched to 86\% in $^{76}$Ge, in the form of about 200 detectors 
in a densely packed array. Each crystal 
will be segmented, and the signals from each segment will be subjected 
to pulse shape analysis.
A half-life sensitivity
is predicted of $4\cdot 10^{27}$~y or 0.02 - 0.07~eV for $m_\nu $ 
after approximately ten years of operation.

The Enriched Xenon Observatory (\EXO ) proposes to use  1-10 tons
of xenon enriched to 60-80 \% in $^{136}$Xe. In contrast to other
proposals, it is planned to discriminate backgrounds by identification 
of the daughter isotope $^{136}$Ba with laser spectroscopic methods. 
If realized successfully, all backgrounds but the $\beta\beta(2\nu)$ 
mode could be suppressed. Two different detector concepts are
under study: high-pressure gas TPC or liquid xenon scintillator.
Sensitivities of $8\cdot 10^{26}$~y, or $0.05 - 0.14$~eV 
for  $m_\nu$, are projected. 

Other interesting projects have been proposed 
including \MOON\,\cite{eji00} (34 t of natural 
molybdenum in a sandwich Mo/scintillator configuration), \CAMEO\,\cite{bel01} 
(1 t of scintillating $^{116}$CdWO$_4$ crystals situated within the 
\borex\ detector), and \COBRA\,\cite{zub01} (CdTe of CdZnTe semiconductors).

In the new $^{76}$Ge initiative at the \LNGS , presented in this Letter of Intent,
we intend  to operate bare germanium diodes enriched to 86\% in $^{76}$Ge
in a high-purity cryogenic medium for shielding against external radiation.
The concept, detailed in the following sections, is based on the observation
that the background signals are largely dominated by external radiation.
By removing most of the cladding and contact 
materials, and immersing the crystals
in an ultra-pure environment, one can operate the diodes largely free
of background. Provided that the background can be reduced to  
10$^{-3}$~\ctsper , 
it will be possible to operate crystals
free of backgrounds up to exposures of 100~(kg$\cdot$y). The experimental
strategy is based on three phases, in each incrementing the target mass.
In Phase I it is planned to operate the existing
almost 20~kg enriched germanium detectors which have been used in the 
\IGEX\ and Heidelberg-Moscow experiment. Conservatively, it is assumed 
that 15 out of almost 20~kg will be operational. About 20 kg of additional 
\geenr\ crystals 
are conceived for Phase II. The third phase has to be defined 
during Phase II and depends on the physics result and the experimental
performance.     
Already after completion of Phase I, it will be possible to 
test the recent claim of evidence for neutrinoless double beta decay.  
\clearpage

\newpage

\section{Design considerations: \\
         background sources and discrimination
         methods}

The main prerequisite for a next generation experiment is a
background reduction by two to three orders of magnitude compared to 
existing experiments. A background index of less
than $10^{-3}$ \ctsper\ is the  goal. This and the
cost of the experiment are the prime design considerations.
In addition, the available space in Gran Sasso Hall A constrains
the design.

While such a tremendous background reduction is difficult to
predict reliably, the contributions of many well identified 
background sources can be extrapolated with Monte Carlo techniques.
Shielding of the germanium diodes with ultra-pure material, such as
liquid nitrogen (LN) or liquid argon (LAr),
is one of the key prerequisites \cite{heu95}.
Following this idea   
the \geni\ \cite{kla99} and \GEM\ \cite{zde01} experiments were proposed.
A more conventional approach with shielding by  copper and lead is
pursued by the \Majorana\ collaboration \cite{maj03}.

\begin{figure}[ht]
\begin{center}
\includegraphics[width=12cm]{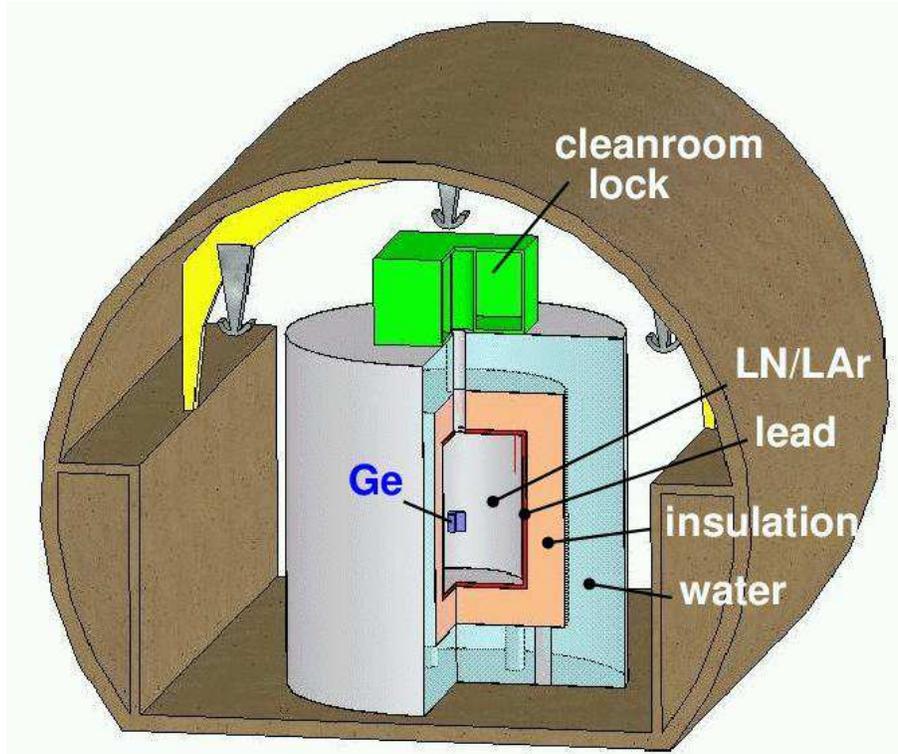}
\end{center}
\caption{\label{fig:generaldesign} 
Schematic drawing of the tank in Gran Sasso Hall A. In the middle of the tank
the Ge diode setup is immersed in the liquid nitrogen or
liquid argon (LN/LAr). Following from the inside
to the outside is a layer of lead,  thermal insulation and a water shield.
The outer diameter is 10 m.}                 
\end{figure}

For the proposed experiment, a combination of the different shielding methods is foreseen.  
The germanium is immersed in liquid nitrogen or argon, but some of the 
shielding is achieved by lead and water.
Fig.~\ref{fig:generaldesign} shows schematically such a  design.
In this section different background sources and vetoing techniques are
discussed.
The various options for the realization of the 
experimental setup will be presented in section~\ref{sec_technicalaspects}.

This section starts with a discussion of the understanding of
the background seen in the Heidelberg-Moscow experiment in the relevant energy
region from 2.0 to 2.1 MeV. Afterwards, the background from external sources
and from sources intrinsic to the
detectors are discussed. Finally, this section ends with a
discussion of the potential
of  scintillation light detection in LAr.


\subsection{Background sources of the Heidelberg-Moscow experiment}
\label{sec:HdMsBackground}
\vskip0.5truecm

A good understanding of the available data is needed to 
estimate the background rejection of the future experiments, 
especially as the existing enriched germanium diodes are used in the
first phase of the experiment. 
The experimental setup of the Heidelberg-Moscow
experiment consisted of 5 copper cryostats, each of them housed an enriched \geenr\ diode.
Four of the detectors were operated in a common lead shield while one detector (no. 4)
was operated in a separate shield of copper and lead \cite{kla03}. 

\begin{figure}[ht]
\begin{center}
\includegraphics[width=12cm]{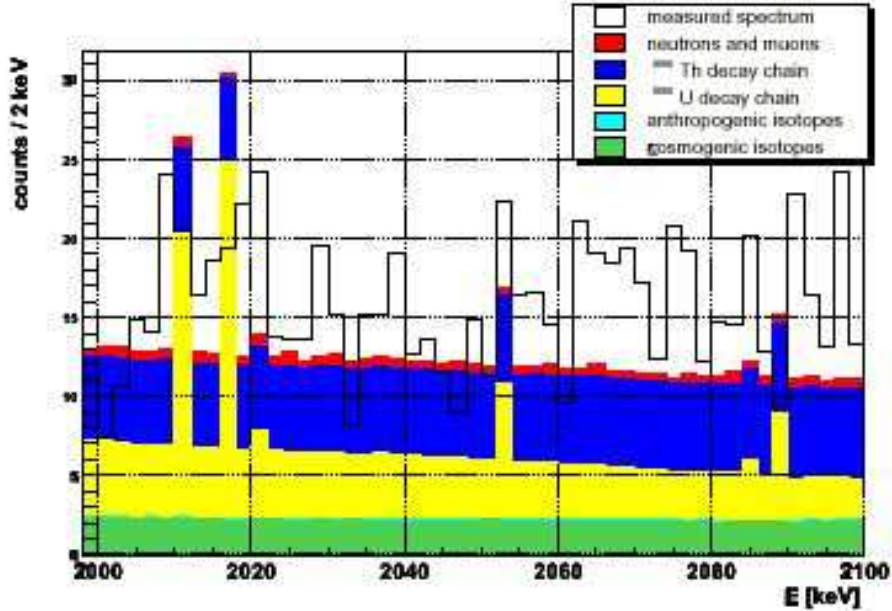}
\end{center}
\caption{\label{heu-1} 
Simulated background of the Heidelberg-Moscow experiment in comparison
with the measured spectrum 
\cite{kla03,dor03}. The background labeled ``cosmogenic isotopes''
refers to $^{60}$Co decays in the copper cryostats. 
}                 
\end{figure}

Fig.~\ref{heu-1} shows the
simulated background spectrum in the energy range from 
2000 to 2100 keV of the 
Heidelberg-Moscow experiment
\cite{kla03,dor03}. 
It identifies the primordial decay chains of U and Th as the main 
contamination sources, with 38\% and 41\%, respectively.
Smaller background contributions originate from the cosmogenic radio-nuclide
$^{60}$Co (16\%), anthropogenic 
contaminations and neutron/muon (in situ) induced events (5\%).

The background contributions are estimated by fitting the 
peak count rates and Compton distributions of simulated spectra
for the individual backgrounds to the total measured spectrum 
(from about 50 to 2700 keV, 49.59 kg y).
In case of a source emitting gamma cascades, like the U/Th chain, the measured energy spectrum
depends sensitively on the distance between source and detector showing e.g. summation peaks of different 
intensities.
This effect has been used to localize the contaminations within the
detector cryostats  and the shielding material.
It is found that the predominant fraction of the U/Th contamination is located at the copper parts 
of the cryostats. 
The same holds also for K contributions, but due to its 
single \gam\ line the 
assignment was more indirect.
The sum of the simulated backgrounds is $660\pm93$ events while
the data spectrum has 803 entries \cite{dor03}. Thus the simulation has
large uncertainties and does not explain the data entirely.

The  cosmogenic activities originate from $^{60}$Co decays and are  
dominantly located in the  copper of the cryostats \cite{die99}.
The upper limits for the fraction of   decays inside the diodes are
between 1.1\% and 7.4\%. Consequently, 
there is no evidence for background sources inside the crystals in the
energy interval of interest.

Table \ref{heu-2} compares the  activities obtained for the copper 
of the individual cryostats  \cite{dor03}
\begin{table}[h]
\caption{\label{heu-2} 
Contaminations of the copper [$\mu$Bq/kg] in the various cryostats of the Heidelberg-Moscow
experiment. The values are deduced from a Monte Carlo simulation of measured spectra \cite{dor03}.
For comparison the corresponding values measured with the GeMPI facility are shown.
} \vspace*{2.mm}
\begin{center}
 \begin{tabular}{lrrr} \hline \hline
 & \radzzs(U)  & \thzza(Th) & \kvn\ \\ \hline
Cryostat of ANG1 & 168$\pm$8  &  84$\pm$7   &   236$\pm$61 \\ 
Cryostat of ANG2 & 91$\pm$4   &  10$\pm$3   &   78$\pm$22 \\ 
Cryostat of ANG3 & 105$\pm$5  &  84$\pm$5   &   927$\pm$46 \\ 
Cryostat of ANG4 & 115$\pm$3  &  87$\pm$4   &   199$\pm$04 \\ 
Cryostat of ANG5 & 100$\pm$4  &  26$\pm$4   &   1632$\pm$49 \\ \hline 
same Cu quality      & $\leq$20 &  $\leq$23   &$  \leq$88 \\
measured with GeMPI  &          &             &           \\ \hline \hline
\end{tabular}
\end{center}
\end{table}
with the directly measured activities of the same 
copper quality by the Ge spectrometer GeMPI \cite{ned00}. 
GeMPI is also operated at the 
LNGS at an even 
lower background level than the Heidelberg-Moscow detectors. Its sample chamber allows to place 
large sample quantities around the detector, for example, 125 kg of Cu in this case.
The strong variation of the activities 
from cryostat to cryostat, which are in most cases 
higher than the upper limits of the bulk measurement with GeMPI, 
suggest that the contaminations are not 
intrinsic to the copper but are rather located on the surface. 
A clear indication of variable surface contamination is given by the 
different $^{210}$Po $\alpha$ peak heights measured in the high 
energy spectra of the five enriched detectors. This contamination which is 
caused by impurities 
on the surface of the p-contacts of 
the crystals varies by a factor of 30 \cite{die99}.

In the liquid nitrogen immersion method the direct cladding material will
be reduced by about 4 orders of magnitude in weight and by about a 
factor of 200 in surface (not counting the crystal surface itself). 
The dominant remaining background will be intrinsic to the diodes.
Since there is no evidence for this contamination, 
a factor of 20 in background reduction is assumed in the calculation of the
physics reach for the first phase of the experiment.

An independent background analysis of the Heidelberg-Moscow data by 
members of the Moscow group is ongoing. 
The method is very similar to the one outlined
above but separates bulk and surface contaminations in the simulations. 
Final  results are expected soon.

\subsection{Background simulations}

Several sources of background have been identified by  
past experiments. These can be classified by the location of
the radioactive isotope: internal to the germanium diode or
external in the material for shielding, in the contacts and in the
concrete/rock of the laboratory.

This section focuses on the simulation of the backgrounds considered
to be most serious.
A more complete list will be given in a later document.
The discussion of the internal background applies to newly
built detectors (Phase II of the experiment) 
while the external backgrounds affect the design of the tank.

For a detailed simulation 
\GEANT 4 version 5.2 patch 2 was used. Only a  simplified setup 
consisting of a tank for liquid nitrogen or argon, 
the liquid  and an ensemble of
27 germanium detectors  was simulated. Each detector has a
height of 78~mm and a diameter of 78~mm which corresponds
to a weight of 2 kg. The  detectors will be ``p-type'' and hence
a dead layer of 0.7~mm at the outside was included in
the simulation. In the middle, a
cylindrical hole of 10~mm diameter and a length of 58~mm is
included for the p contact.
The crystals are arranged in a $3\times 3\times 3$ array
with a spacing of 12~mm.

In addition a CPU time optimized Monte Carlo
for extensive simulations of gamma interactions is used. This program
simulates Compton scattering, pair production  and photo electric
effect and performs about a factor of 30 faster than GEANT4.
However, only geometries with a single diode have been implemented.
Simulations were carried out for internal as well as for external background.

\subsubsection{The internal background from cosmogenic isotopes}
\label{sec:intrinsicbkg}

During storage at sea level the germanium is exposed to hadronic
radiation, especially neutrons. These cause spallation in the germanium
and hence a variety of radioactive isotopes are produced. These processes
can be simulated and past experience shows that the results of 
simulations agree with 
measurements within a factor of two \cite{avi92,mai96}.

Most relevant for the neutrinoless double beta decay are the decays 
of $^{68}$Ge and $^{60}$Co
since  $Q$ values above $Q_{\beta \beta} =$ 2039 MeV 
occur in the decay chain, and 
the lifetimes are in the range of years.

\vspace*{2.mm}
\noindent {\it Cosmogenic $^{60}$Co background in the germanium diode}
\vspace*{2.mm}
 
\noindent The cosmogenic production of $^{60}$Co in a germanium detector
is about 4 atoms/(kg d) 
at sea level \cite{avi92}.
Therefore an exposure of one day corresponds  to an activity of 
$4 \, {\rm ln} (2) / T_{1/2} = 0.017 \, \mu$Bq/kg. 
This number agrees within a factor
of 1.6 with the
value from reference \cite{bau99}. \footnote{The value given in
reference \cite{bau99} of 
0.18 $\mu$Bq/kg after a 10 day exposure and 3 years of storage underground
corresponds to 0.027  $\mu$Bq/kg for a one day exposure without underground
decay time.}

\begin{figure}[ht]
\begin{center}
\vspace*{-10.mm}
\includegraphics[width=\textwidth]{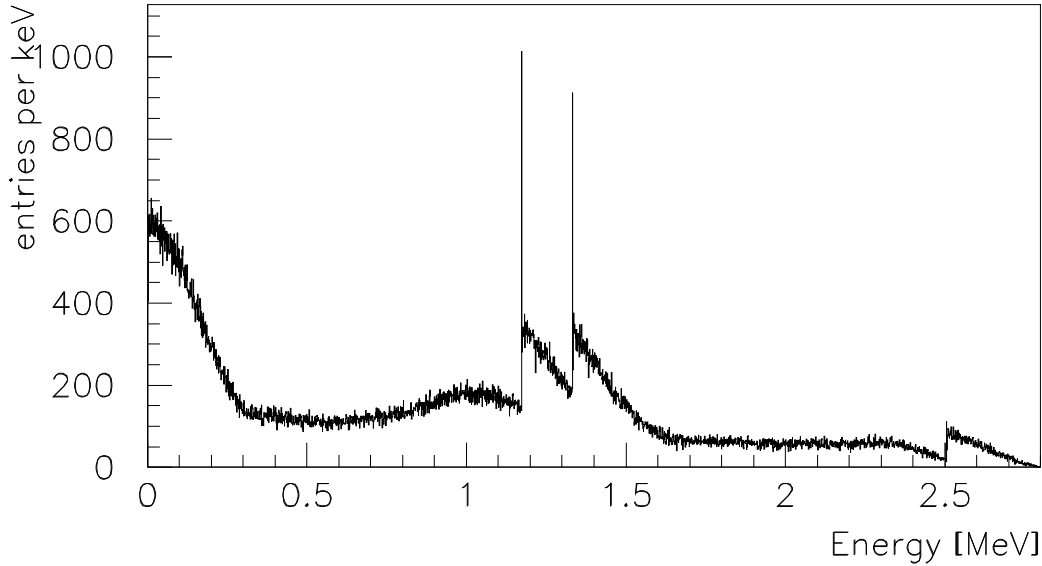}
\vspace*{-20.mm}
\end{center}
\caption{\label{fig:co60} Energy deposition in a germanium detector
from $^{60}$Co decays located inside the crystal.
The decay chain starts with a $\beta^-$ emission of an endpoint energy of 300~keV
followed by two prompt $\gamma$ quanta of 1.173 and 1.332~MeV. 
}                 
\end{figure}

If the detectors are fabricated underground, an exposure time
of 10 days after  zone refinement may be realistic and shall be used
here. Since the zone refinement will largely remove the $^{60}$Co atoms,
the exposure time results in an activity of $0.17 \mu$Bq/kg 
which corresponds to 5.4 decays/(kg y).
Fig.~\ref{fig:co60} shows the simulated energy deposited inside the
diode. In  one out of 6000 decays the deposited energy is at
$Q_{\beta \beta}$ within a 1 keV window.
The resulting background index is therefore $0.9\cdot 10^{-3}$ \ctsper .
Most of these events  can be rejected by requiring an 
anti-coincidence with other detectors and for segmented detectors
by an anti-coincidence of the segments.
If the decay occurs in the inner crystal of the simulated setup
and if the deposited  energy is close to $Q_{\beta \beta}$, 
in about 85\% of the events some energy is deposited in a second
crystal.
For a detector at the corner about half of the events have energy
deposition in a  second crystal.

Another proposed background rejection method  is the 
segmentation of the detector contact
which allows for a better localization of the energy deposition. 
Since electrons  will deposit their energy very localized, the
anti-coincidence of segments can be applied.
For a  four-fold segmentation
along the axis of the diode  (axial segmentation)
an additional suppression of about 5 can be expected. 
Slightly worse numbers apply for a 
four-fold azimuthal
segmentation (pie slices). In total a factor of 30 in background
suppression can be achieved (for the center crystal) 
by the anti-coincidence method.

Pulse shape analysis provides an additional mean to suppress background.
 This has not been studied so far
but is expected to result in an additional rejection factor
of more than 2. Beside the methods used in previous Ge experiments or
proposed for the \Majorana\ experiment
\cite{maj03,kla04}, there is also the possibility to  extend the
method used by the \GNO\ experiment for the discrimination of 
multiple site events in proportional counters \cite{pan04}. 

If LAr is used as the cryogenic fluid, the detection of scintillation
light would be a powerful discrimination method. According to simulations
(see section~\ref{sec:arscintillation}) the $^{60}$Co background at 
$Q_{\beta \beta}$ can be suppressed by a factor of 100.  The ultimate
limit is given by the thickness of the dead layer at the outer surface of 
the detector. 

\vspace*{2.mm}
\noindent {\it Cosmogenic $^{68}$Ge background in the germanium diode}
\vspace*{2.mm}

\noindent According to calculations in reference \cite{avi92}, the production
rate of $^{68}$Ge in $^{76}$Ge is about 1 atom/(kg d). Here the time
since the last isotope separation step  is relevant. Even if
the detector is fabricated underground the activation time could
be several months. Since $T_{1/2} \sim 270$~d, the activation will
be a large fraction of the saturation activity. 
Here we will assume 180 days for the time
between the isotope separation and the storage underground.
This results in 40\% of the saturation activity.

\begin{figure}[ht]
\begin{center}
\vspace*{-10.mm}
\includegraphics[width=\textwidth]{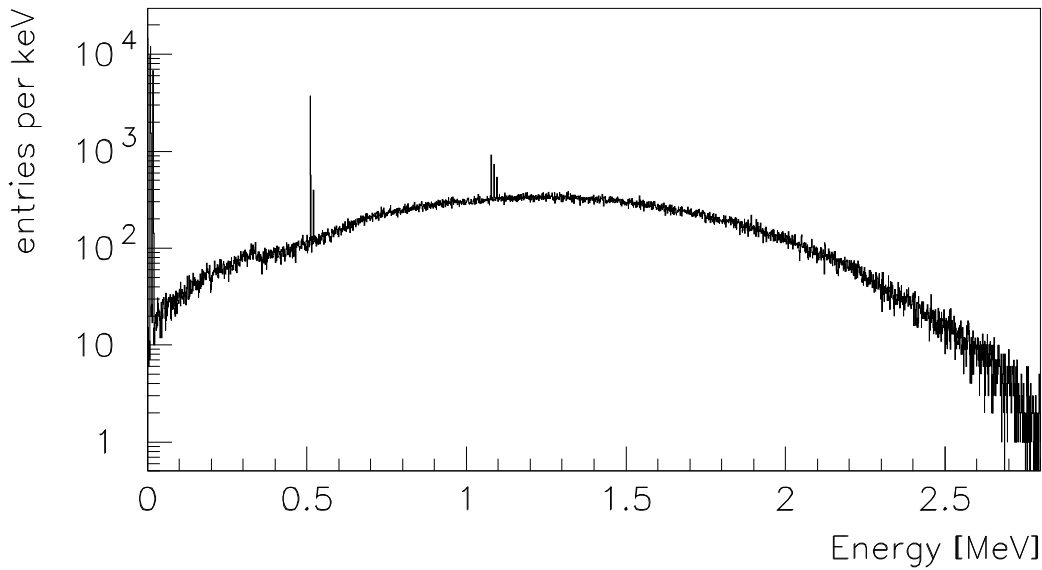}
\vspace*{-20.mm}
\end{center}
\caption{\label{fig:ge68} Energy deposition in a germanium detector
from $^{68}$Ga decays inside the crystal. 
}                 
\end{figure}

The saturation activity corresponds to 1 decay/(kg d)  at the time
the detector is brought underground. This corresponds to 
400 $^{68}$Ge atoms/kg, and
60\% of these will decay in the first year. 
Fig.~\ref{fig:ge68} shows the spectrum of deposited energy. In about one
out of 5000 decays the energy is within 1~keV of $Q_{\beta \beta}$.
This yields a background of $19\cdot10^{-3}$ \ctsper\
which is obviously not acceptable.
Several methods exist to reduce this background:
\begin{itemize}
\item Waiting: the small half life makes waiting an option.
      After three years the activity is 1/16.
\item Anti-coincidence of detectors: 
      the $\beta^+$ decay of $^{68}$Ga yields two 511 keV photons from 
      positron annihilation. These will
      often deposit energy in neighboring detector. For the central 
      detector the anti-coincidence would result in a reduction by a factor
      2.5. For a decay in an outer detector the background is reduced by
      a factor of 1.4.
\item Anti-coincidence of segments: 
      the four-fold axial segmentation yields an additional suppression
      factor of 4.5.
\item Anti-coincidence with scintillation light in case the cryogenic
      fluid is LAr.
\item Tagging the $^{68}$Ge decay to $^{68}$Ga and vetoing the subsequent $^{68}$Ga decay: 
      About 86\% of the
      $^{68}$Ge decays occur via electron capture from the K shell which
      results in a $\sim$10 keV
      energy deposition when the empty K shell location is filled. 
      The $\beta^+$ decay of $^{68}$Ga follows with a half 
      lifetime of $T_{1/2} =$ 68 min. The time correlation of the two
      decays is therefore  a powerful rejection tool. 
      Consequently, a factor of 5 in background rejection is feasible.
\end{itemize}

Using these rejection factors,  the background index in the first 
year (for the central
crystal) can be reduced to $0.3\cdot10^{-3}$ \ctsper\ 
and falls exponentially with $T_{1/2} \sim$ 270 days.

As before, the background discrimination potential 
from the pulse shape analysis has not been explored yet.

\subsubsection{External background}
\label{sec:externalbkg}

The external background consists of photons from  primordial decay chains,
neutrons and muon induced background. 
So far, most of our emphasis is focused  on
the  suppression of the 2.615 MeV photon from the $^{208}$Tl decay. This
background influences the design of the tank considerably.

\vspace*{2.mm}
\noindent {\it $^{208}$Tl background}
\vspace*{2.mm}

\noindent The activity in the concrete and the rock in \LNGS\ from $^{208}$Tl  
results in a flux of about $1.4\cdot10^7$ decays/(d m$^2$) \cite{arp92}. 
For a cylindrical tank with
8 m inner diameter and 8 m height this corresponds to a total flux of
$1.5\cdot 10^{12}$  photons per year.

The simulation of such a flux with \GEANT 4  is too CPU intensive. There are
three alternatives to estimate the background. 
With \GEANT 4 one can simulate smaller tanks of
different size and
then extrapolate to the background for the full size. 
Alternatively the fast simulation
program mentioned above was used.
The third method is based on the peak  
to Compton  ratio and the detection efficiency
as measured with  a $^{228}$Th source for a similar detector.\footnote{The
peak to Compton ratio is defined here as the number of events in
the 2.615 MeV peak to the number of events in the
energy interval of 2.00 - 2.08 MeV.}
Then the  absorption length of the $^{208}$Tl photon
is used to analytically calculate the background index for a given shielding
and surface activity.

For example, the shielding of a surface activity of 0.26 Bq/cm$^2$ from the
concrete/rock with 570 cm of liquid nitrogen results in a background
index 
\begin{eqnarray}
B & = & 5\cdot10^{-3}\frac{cm^2}{keV \cdot kg} \,  0.26 
                          \frac{1}{cm^2 \cdot sec}
    \, {\rm e}^{-570\cdot 0.03115} 
  = 2.5 \cdot 10^{-11} \frac{cts}{keV \cdot kg \cdot sec} \\ 
  & = & 0.8 \cdot  10^{-3} \frac{cts}{keV \cdot kg \cdot y}
\end{eqnarray}
where $5\cdot10^{-3}$cm$^2$/(keV$\cdot$kg) is the measurement result
and $\mu = 0.03115$~cm$^{-1}$ is the absorption coefficient in liquid nitrogen.

All methods predict within a factor of two the same background
suppression. 
Different design options for the tank  are discussed
in section~\ref{sec_technicalaspects}. 
The resulting background indices are calculated
with the above formula.

Note that the detection of scintillation light in LAr
could result in a background rejection factor of 20 (see 
section~\ref{sec:arscintillation}). This has not been taken
into account in the current design.


\vspace*{2.mm}
\noindent {\it Neutron induced background}
\vspace*{2.mm}

\noindent The main sources of neutrons in the LNGS are from spontaneous 
fission (dominated
by the $^{238}$U isotope)  and from ($\alpha$,n) reactions in the concrete 
and the rock. The maximum neutron energy from these processes is about 9 MeV.
Neutrons above this energy originate from muon interactions which
have a much smaller flux.

\begin{figure}[ht]
\begin{center}
\vspace*{-10.mm}
\includegraphics[width=\textwidth]{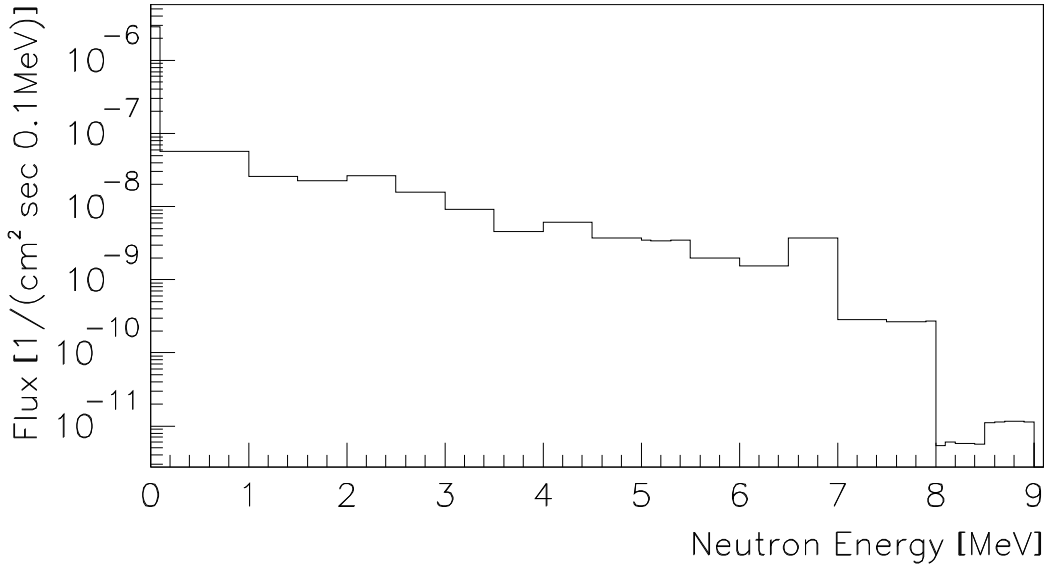}
\vspace*{-20.mm}
\end{center}
\caption{\label{fig:neutronspectrum} Prediction of the energy spectrum of neutrons
in LNGS Hall A from fission and ($\alpha$,n) reactions \cite{wul03}. 
}                 
\end{figure}

No detailed measurements of the neutron energy spectrum exist. Instead the
spectral shape is taken from a simulation of the above processes 
\cite{wul03}. The predicted integral  flux of 
$4\cdot 10^{-6}$ neutrons/(cm$^2 \cdot $sec)  is in agreement with
measurements.\footnote{For a more detailed discussion see \cite{wul03}
and references therein.}
The energy spectrum  is shown in Fig.~\ref{fig:neutronspectrum}.
The peak in the spectrum at 6.75 MeV is due to ($\alpha$,n) reactions
on magnesium and carbon. Neutrons above  7 MeV are from fission.

A spherical vessel of 7 m diameter plus a variable
thickness of polyethylene as a moderator is simulated. 
The results shown here will not
depend on the exact shape of the tank. It should be noted, however, that
the simulation assumes hermetic moderator shielding.

The simulation in GEANT4 is based on  cross section tables
for all the relevant isotopes.
Tables are available for elastic scattering,  capture  (n,$\gamma$)
and inelastic reactions
(n,n$\gamma$), (n,p), (n,p$\gamma$).

In Table~\ref{tab:neutron1} the neutron flux and the
mean neutron energy after the polyethylene moderator
is given. With increasing thickness more and more slow neutrons
are captured, and consequently the spectrum becomes harder. 
Since nitrogen also acts as an additional moderator, neutron background
is more critical for argon filling, and we primarily address this case.
In liquid argon
neutrons loose typically only a few percent of their energy 
per elastic scattering
and the mean free path is  15-20 cm above a  kinetic energy of 0.5~MeV. 
Consequently most neutrons are backscattered and
stopped in the moderator. 
\begin{table}[ht]
\caption{\label{tab:neutron1} Relative neutron flux $\Phi$ and average 
neutron energy $E$ after a polyethylene moderator for
different moderator thicknesses.} \vspace*{2.mm}
\begin{center}
 \begin{tabular}{lcccccc} \hline \hline
 thickness  & 0 cm & 10 cm & 20 cm & 30 cm & 40 cm & 50 cm \\ \hline
$\Phi$   &  1   & 0.22  & 0.031 & $6\cdot 10^{-3}$ & $1.3\cdot 10^{-3}$ &
                                     $3.8\cdot 10^{-4}$ \\ 
$E$ [MeV] & 0.52   & 0.61    & 0.81 & 0.97 & 1.09 & 1.16    \\ \hline \hline
\end{tabular}
\end{center}
\end{table}

\begin{figure}[ht]
\begin{center}
\vspace*{-15.mm}
\includegraphics[width=\textwidth]{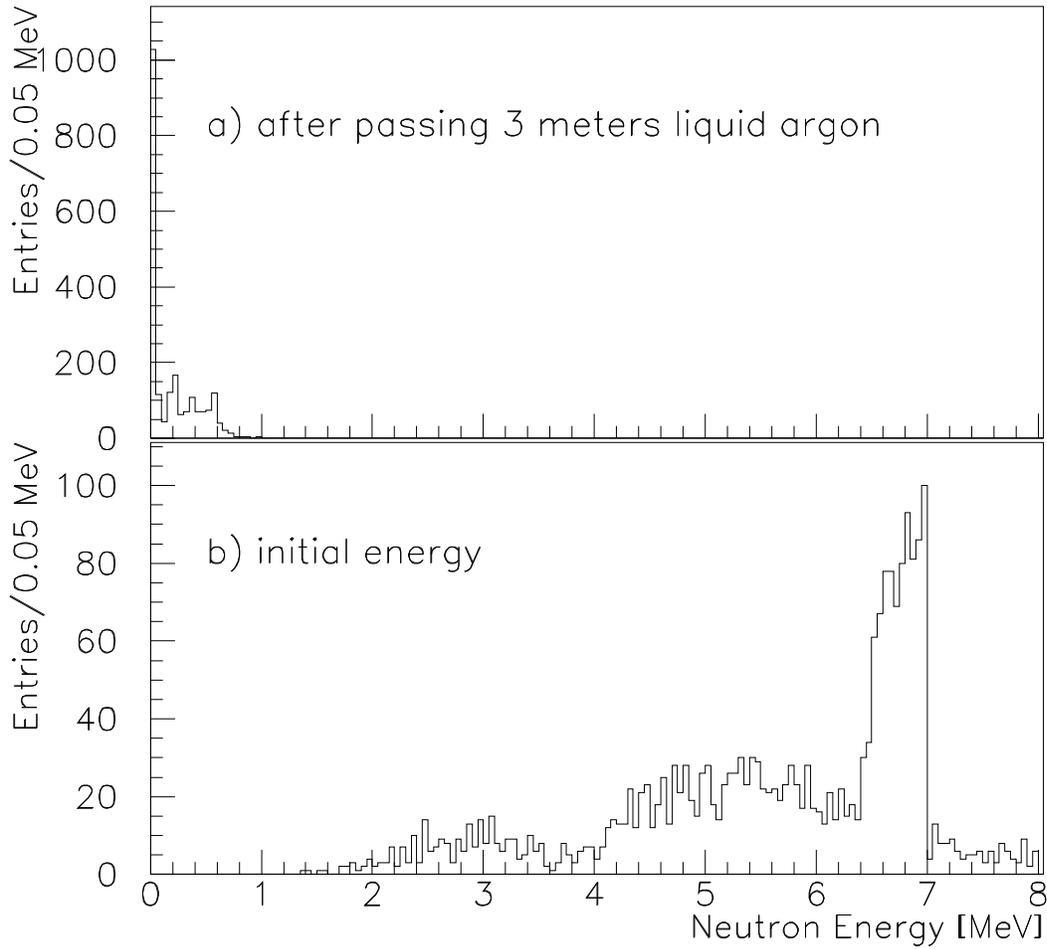}
\vspace*{-20.mm}
\end{center}
\caption{\label{fig:neutron50cm} a) Neutron energy after 40 cm of moderator
and 3 m of liquid argon. b) Initial energy of these neutrons.
}                 
\end{figure}

A  moderator thickness of 40~cm was simulated in detail.
In Fig.~\ref{fig:neutron50cm}~a)  the neutron energy spectrum is 
shown for
those neutrons that pass the moderator and 3 m of liquid argon. Part b) shows
the initial energy of the same neutrons. 
All slower neutrons are captured and only the neutrons with large
initial energy reach
the germanium diodes. Here elastic scattering dominates again and
almost no neutron is captured in the germanium. 
The Monte Carlo sample size corresponds to twice the integrated flux 
for one year. No event was observed in the energy region of $Q_{\beta \beta}$.
Hence a moderator thickness of 30~cm resulting in a factor of 5 higher
flux is also  sufficient.

For nitrogen the neutron background is even less of a problem, since
the energy loss per scattering is almost a factor of 3 larger and the cross
sections are typically a factor of 10 larger.
Water will provide a similar shielding as polyethylene.

Presently all options include a thick layer of water ($>$50~cm) as shield. 
This will ensure that the neutron background is negligible.

\vspace*{2.mm}
\noindent {\it Muon induced backgrounds}
\vspace*{2.mm}

\noindent The outer area of the tank is rather large. An instrumented 
layer of water
for the detection 
Cherenkov light from muons is  therefore a cost-effective solution for
a muon veto. 
No detailed design or simulation of the efficiency
of such a detector is available at the moment.

For the \geni\ project already a 96\% efficiency of the veto resulted in a background
index of $2\cdot 10^{-6} $ \ctsper\ \cite{bau99}, and therefore details of the veto
design are not considered critical at this time.
This veto will also  reject most of the inelastic muon reactions
such as the production of secondary neutrons in the lead shield or the
LN/LAr.


\vspace*{2.mm}
\noindent {\it Background from contaminations of the liquid nitrogen~/~argon} 
\vspace*{2.mm}

The \borex\ collaboration has measured upper limits for the contamination
of their liquid scintillator of less than $3.5\cdot 10^{-16}$ g/g for $^{238}$U
and less than $4.4\cdot 10^{-16}$ g/g for $^{232}$Th \cite{ali98}. 
Similar limits are expected for LN/LAr since they are also produced
by fractional distillation. These limits 
correspond to 32 $^{238}$U decays and 11 $^{232}$Th decays 
per m$^3$ and year for liquid nitrogen.

Due to heat losses some of the liquid  will evaporate and will be lost.
An upper limit of 200 $\mu$Bq/m$^3$ for the $^{222}$Rn
contamination is assumed for the liquid supply \cite{mpi03}. This corresponds
to 3.2 atoms/m$^3$. During one year, 200 m$^3$ will be exchanged
(once the entire volume) and thus  640  decays have to be taken into account.
Alternatively, if the contamination does not decay
with $T_{1/2} = $ 3.8 d  but stays constant (due to permanent
emanation), 6300 decays/m$^3$  have
to be taken into account per year. A simulation shows that even this
very conservative assumption leads to an upper limit  well
below 10$^{-3}$ \ctsper . This can be further reduced by 
anti-coincidence methods and pulse shape analysis.

%
%


If the choice of the liquid is  argon, then additional background
from $^{42}$Ar decays have to be taken into account. 
The upper limit on the activity from $^{42}$Ar is 
40 $\mu$Bq/kg  \cite{ash03}. A simulation shows that this 
results in a background of  $4\cdot 10^{-5}$ \ctsper .
Beta decays of $^{39}$Ar have a $Q$ value of 0.6 MeV and do not
contribute to the background for neutrinoless double beta decay.

\vspace*{2.mm}
\noindent {\it Background from the detector surface and  the holder material} 
\vspace*{2.mm}

The experience of many low background experiments shows that the surface
contamination is by far larger than the bulk activity. 
For  coaxial detectors
the inner well is critical. To estimate
the sensitivity of the experiment to this background, 
$^{208}$Tl decays were generated at this location. 
For 2.5\% 
of these decays, the energy  deposition is in the interval 2.0-2.08 MeV in
one of the crystals (out of 27).
About 90\% of these events can be rejected by the anti-coincidence with
a second detector and an additional factor of 4 can be gained by a 4-fold
axial segmentation.
Consequently, for a background index of $10^{-3}$ \ctsper\ 
the surface activity of $^{208}$Tl
has to be below 100 $\mu$Bq/m$^2$   without any
anti-coincidence method ($\sim$ 6 decays per year and detector) 
and below 4 mBq/m$^2$ with the above mentioned
rejection factors. The background contribution from $^{214}$Bi decays is
about a factor of 10 smaller for the same activities.

For the detectors of the Heidelberg-Moscow experiment there is clear
evidence for surface contamination for two of the five detectors 
\cite{die99,bak03}. The level is of the order of 45 decays of 
$^{210}$Pb per year and detector. For the $^{232}$Th
decay chain the intensity is approximately a factor of 4  smaller
and consequently one would expect about 4 $^{208}$Tl decays on the detector
surfaces per year. The resulting background index is 
$\le 10^{-3}$ \ctsper . For the other detectors the contamination
level of $^{210}$Pb is about a factor of 10 smaller or not seen at all.

Concerning the energy deposition in the
crystals and the fraction of events with energy depositions in neighboring
diodes or different segments, 
contaminations in the holder material of the diodes have similar effects as  
surface contaminations.
Therefore for 10 gram holder weights per diode and
a background index of  $10^{-3}$ \ctsper , the contamination
of the material should be below 20 $\mu$Bq/kg of $^{208}$Tl 
($1.5\cdot 10^{-11}$
g/g of $^{232}$Th).\footnote{The background from $^{214}$Bi decays is  
about a factor
of 10 smaller for the same contamination level.} 
If the anti-coincidence methods are applied this number can be larger
by up to a factor of 40. For the acryl material used by SNO,
a contamination level of $10^{-12}$ g/g for $^{232}$Th was found,  
resulting in a background index of 
$\le 10^{-4}$ \ctsper .

\subsection{Background summary}
\label{sec:bkgsummary}

{\bf Phase I:} The external background will be reduced to a
level of $10^{-3}$ \ctsper .
From the analysis of the Heidelberg-Moscow data we are confident
that the overall background can be reduced by at least a factor of 20 once the
diodes with contacts of reduced mass are immersed in LN/LAr 
(see section~\ref{sec:HdMsBackground}).
The background index is then   $\le 10^{-2}$ \ctsper .

\phantom{in Table\~ref{tab:bkgsummary}.}
\begin{table}[h]
\caption{\label{tab:bkgsummary} 
Summary of background estimates for Phase II of the experiment. For
the assumptions see text. The external $\gamma$
background is for shielding with LN.
} \vspace*{2.mm}
\begin{center}
 \begin{tabular}{lc} \hline \hline
source & background index in $10^{-3}$ \ctsper \\ \hline
external $\gamma$ from $^{232}$Th, $^{228}$U &  1    \\
external neutrons                            &  $\le 0.05$ \\
external muons                               &   $\le 0.05$ \\
internal $^{68}$Ge                           &  0.9 \\
internal $^{60}$Co                           &  0.1 \\
contamination in LN/LAr ($^{222}$Rn)         &  $\le 0.1$ \\
contamination in holder material             &  $\le 0.1$ \\ 
surface contamination                        &  $\le 0.1$ \\ \hline \hline
\end{tabular}
\end{center}
\end{table}
{\bf Phase II:} A summary of the estimated background contributions is given
in Table~\ref{tab:bkgsummary}.
For the estimate of the internal background,
realistic assumptions are made 
for the fabrication times: 
The time between the enrichment process and the storage underground is
180 days (relevant for $^{68}$Ge production),
the time between the last zone refinement and the storage 
underground is 30 days (relevant for $^{60}$Co production) and the
time between the storage underground and
the start of the measurement is 180 days (relevant for $^{68}$Ge decay).
For the anti-coincidence methods a four-fold
axial segmentation and a corner location of the diode in the setup is used.
In addition, a factor of two for background rejection is assumed from
pulse shape analysis. 

\subsection{Performance with instrumented shield for LAr}
\label{sec:arscintillation}

The  design of our proposal consists of germanium diodes immersed
in liquid nitrogen or argon. The liquid serves as a high purity {\it passive} 
shield  against radiation. Ionizing radiation that creates 
background signals in the diodes with energies close to  
$Q_{\beta\beta}$ typically has energies greater 
than the one deposited in the germanium crystals. 
Part of this
energy is dissipated in the shielding liquid and is `invisible'.
An option under study is to instrument the shielding medium 
and to measure the energy deposition which can be used as 
an anti-coincidence signal. 
Liquid nitrogen provides only weak signals from 
scintillation and Cherenkov light emission.
The scintillation properties of liquid argon are well established,
see \cite{kub79, dok90, hit83, cen99}: 
about 40,000 photons are emitted per MeV of deposited energy.
This is approximately four times the number observed in organic 
liquid scintillators. Photons are emitted in the de-excitation 
of the Ar$^*_2$ excimer with a wavelength of 128 nm.
Decays from singlet and triplet excited states give rise to 
a fast 6~ns and a slow 1.6~$\mu$sec component 
with an intensity ratio of 0.3  for excitation 
with electrons, and of 1.3 for alpha particles \cite{kub79}.

In order to use the scintillation light  in anti-coincidence
with the germanium diodes, one has to detect the 128~nm scintillation
photons with high efficiency. 
The  use of wavelength shifting materials to move the photon wavelength
into the region of maximal sensitivity of Bi-alkali photomultipliers
is under investigation.
Technical details of this developments are summarized in 
section~\ref{sec:RandD}. 
A Monte-Carlo simulation of a 
2~kg diode immersed in liquid argon demonstrates the potential power
of the method.   
We assume that a threshold of 100~keV can be achieved for the detection of scintillation
light. 
\begin{figure}[ht]
\vspace*{-1.cm}
\begin{center}
\includegraphics[width=12cm]{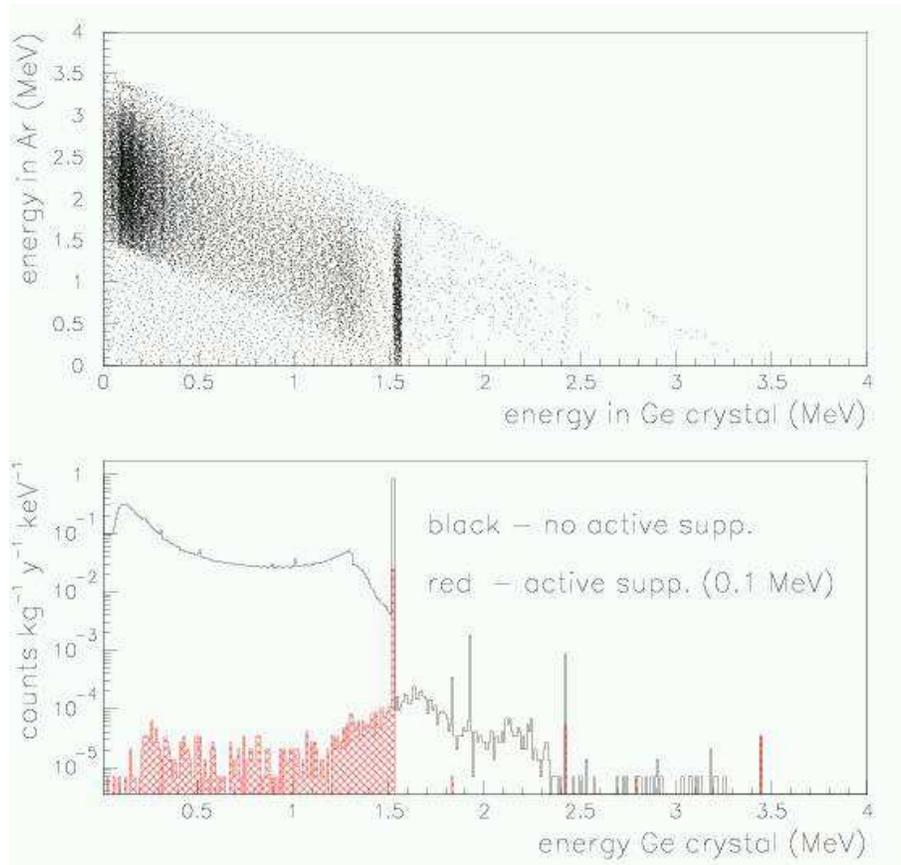}
\end{center}
\vspace*{-1cm}
\caption{\label{fig:Ar42lar} 
Suppression of external $^{42}$K ($^{42}$Ar progeny): 
Scatter plot of the energies deposited in the liquid argon and the Ge diode (top), and
the energy spectrum seen by  the 
germanium crystal with and without anti-coincidence 
assuming a threshold of 100~keV (bottom). }
\end{figure}
Figure~\ref{fig:Ar42lar} displays the results of simulations of 
$^{42}$K decays, the progeny of $^{42}$Ar ($Q_\beta = 0.6$~MeV, 
$t_{1/2}=33$~y),  
homogeneously distributed in the liquid argon.
$^{42}$K has a maximum electron energy of 3.5~MeV and a
weak $\gamma$ line at 2.424 MeV and thus 
is a relevant background for double beta decay. 
A ratio of $^{42}$Ar/$^{nat}$Ar of 
$3\cdot 10^{-21}$ has been assumed \cite{bar02} for the 
simulation. A background suppression of more than a factor of 100
is achieved in the $Q_{\beta\beta}$ region.
  
\begin{figure}[ht]
\vspace*{-1.cm}
\begin{center}
\includegraphics[width=12cm]{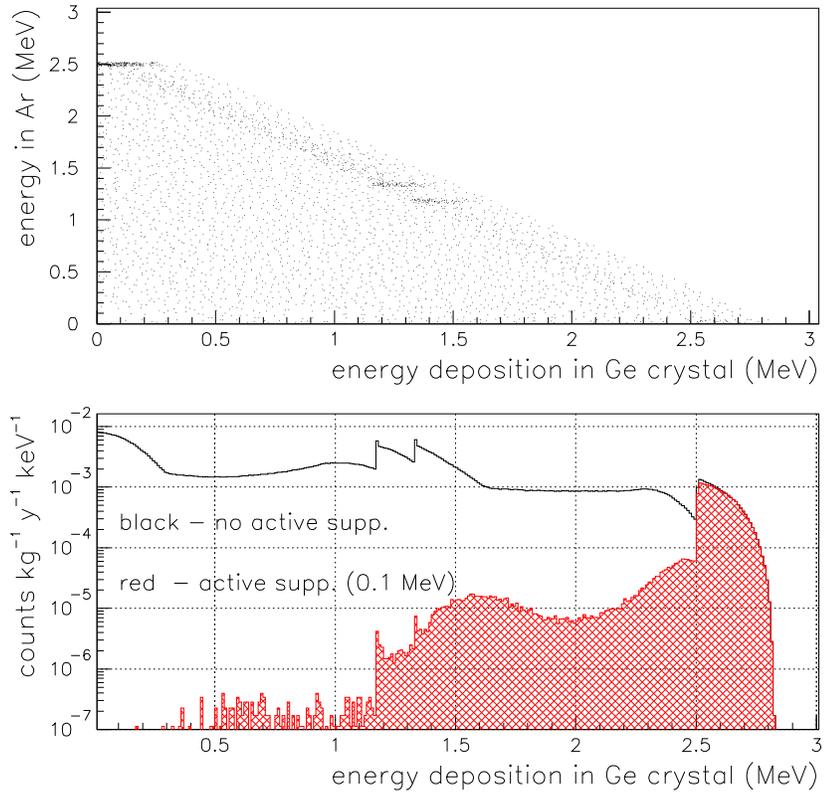}
\end{center}
\vspace*{-1cm}
\caption{\label{fig:co60lar} 
Suppression of internal $^{60}$Co activity: 
Scatter plot of the energies deposited in the liquid argon and the Ge diode (top),
and the energy spectrum seen by  the 
germanium crystal with and without anti-coincidence 
assuming a threshold of 100~keV (bottom). }
\end{figure}

Figure~\ref{fig:co60lar} shows a simulation of cosmogenic 
$^{60}$Co decays which are placed homogeneously inside 
 a germanium crystal. An activity of 0.18~$\mu$Bq/kg is assumed.
Again, one gains two orders of magnitude in background suppression.

 
%
%
\clearpage

\newpage
\section{Physics reach}

The experiment will have several phases. In the first phase 
the existing enriched germanium detectors will be used while new
diodes need to be fabricated for the second phase. The 
sensitivities are discussed in the following.

\subsection{Phase I}

For the case that no background events are observed,
the 90\% confidence limit (C.L.) on the half lifetime for the neutrinoless 
double beta decay is given by 

\begin{equation}
  T_{1/2} > 2.4 \cdot 10^{24}   \cdot \epsilon \cdot a \cdot M \cdot t \,~~~[y]
\end{equation}

\noindent with $\epsilon$ being the 
detection efficiency, $a$ being the enrichment
fraction of the $^{76}$Ge isotope, 
$M$ being the total active mass of the diodes in kilograms and $t$
being the measurement time in years.

If the number of background events is large and  Gaussian errors can
be assumed,  the same confidence level limit is given by

\begin{equation}
  T_{1/2} > 4.3 \cdot 10^{24}   \cdot \epsilon \cdot a 
                \sqrt{ \frac{M \cdot t}{B \cdot \Delta E} } \,~~~[y]
\end{equation}

\noindent with $B$ being the background index in \ctsper\ and
$\Delta E$ being the energy resolution in keV.

For the first phase of the experiment only  existing enriched
detectors will be used. 
In total, the active mass of the existing enriched detector is almost
20~kg.
Here we will only assume an active mass of 15~kg, since
some the detectors might not work reliably.

The reduction of the external backgrounds to a level of less than $10^{-3}$ 
\ctsper\ is ensured by the design of the tank and the reduction and
selection of the detector mounting material. 
For the intrinsic  background we assume a level of $10^{-2}$ \ctsper\
as discussed in section~\ref{sec:HdMsBackground}.

Assuming one year of data taking and a resolution of 3.6~keV,
the expected number of background events is 0.5 counts. 
If no  event is observed (60\% chance),
a 90\% C.L.~of $T_{1/2} > 3.0\cdot 10^{25}$~y can be established for
a detection efficiency of $\sim 95\%$ and an enrichment fraction of $86\%$.
This results in
an upper limit on the effective neutrino mass of $m_{\nu} < 0.24-0.77$~eV.
The mass range is determined by the matrix elements quoted in \cite{ell02}.
If the possibility of a non-zero event count is taken into account
(with  weights determined by a poison distribution of mean 0.5 events)
then the  upper limit becomes $T_{1/2} > 2.2\cdot 10^{25}$~y or,
translated into an effective neutrino mass, $m_{\nu} < 0.28-0.9$~eV. 

%
%

The current claim of a signal
for neutrinoless double beta decay \cite{kla04} is based on
an excess of $28.8\pm6.9$ events
for a total statistics of 71.7 kg$\cdot$y.
With a statistics of one year data taking of
15 kg$\cdot$y and a similar detection 
efficiency we would observe $6.0\pm1.4$ events
above a background of 0.5 counts. If no event is observed, the
claim would be ruled out with 99.6\% confidence level. If one event were observed,
it would be a 97.8\% confidence level. However, if six or more
events would be observed, this would correspond to a 5 sigma confirmation.

If some of the detectors exhibit a larger background index the same
sensitivity can be reached with a subset and a correspondingly larger
running time.

\subsection{Phase II}

In the second phase of the experiment new 
germanium detectors will be added to
the setup. A background index  
of $10^{-3}$ \ctsper\ is achievable, if the detector exposure
can be kept to the limits discussed in section~\ref{sec:bkgsummary}.

It is foreseen to accumulate  statistics of about 100 kg$\cdot$y 
within 2-3 years. 
With a resolution of 3.6~keV, 0.36 background counts are expected.
Consequently there is a large chance that no event is observed. 
The limit on
$T_{1/2}$ would improve by a factor of six to $2\cdot 10^{26}$~y.
This translates to an upper limit of the effective neutrino mass
of 0.09~-~0.29~eV, depending on the nuclear matrix element  used. 

\newpage

\section{Technical aspects of the experiment}
\label{sec_technicalaspects}

\subsection{Overview}
\begin{figure}[h]
\begin{center}
\includegraphics[width=13cm]{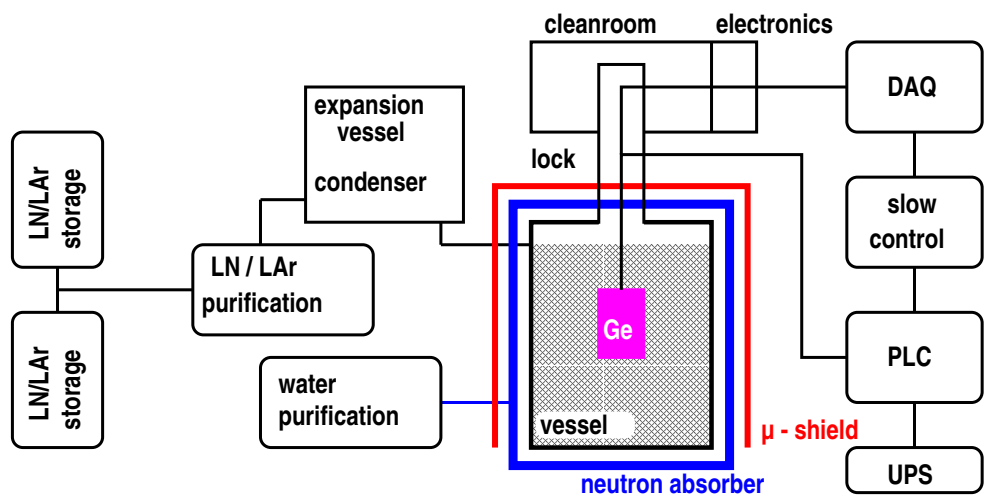}
\end{center}
\caption{\label{exp-schem} 
Schematic of the experimental setup. Part of the muon shield is integrated in a water tank
serving both as shield against $\gamma$ background and neutron absorber. 
}		  
\end{figure}

Figure~\ref{exp-schem} shows a schematic of the experimental setup. Via the cleanroom, the Ge diodes 
are brought via the cleanroom to the lock through which they are installed in the
cryogenic vessel filled with either liquid nitrogen (\ln ) 
or liquid  argon (\lar ). The cryofluid serves both as cooling and shielding medium for the diodes.
The detector suspension and contacts are made out of a minimum of carefully screened materials of
ultra-high radiopurity. 
The vessel is hermetically enclosed by a passive neutron absorber consisting of water and/or
polyethylene. Signals induced by incident muons are rejected with the help of muon detectors -
scintillators on top of the vessel and photomultiplier tubes in the water detecting Cherenkov light.

The signals of the Ge diodes are led via feed-throughs into the electronic cabin on top of the vessel
where their pulse shapes are digitized in flash ADCs. The PC based data acquisition system allows 
the processing, visualization, quality control, and storage of the incoming data. 
The slow control supervises all auxiliary systems including the low and high voltage power 
supplies and the monitoring of the parameters characterizing the status of the 
experiment like temperature, pressure,
detector currents, etc. 
The programmable logic controller (PLC) is an autonomous system of high
reliability for the monitoring of safety relevant parameters like vessel pressure, oxygen concentration,
and temperature. An uninterruptible power supply (UPS) allows the system to continue data taking or 
to reach a safe state in case of an electrical power outage.    

The cryogenic vessel is sealed with metal gaskets and operated at 0.2~bar overpressure to prevent 
contamination of the
cryofluid by radon from the environment. It might be necessary to purge the cryofluid continuously
from Rn contamination. This is done by low temperature charcoal adsorber columns which are 
used also to supply fresh Rn free cryofluid as substitution for evaporated \ln/\lar . 
In case of \lar\ it might be, however, more economic to re-condense  the \lar\ vapor with
either a cryogenerator or an over-pressurized \ln\ radiator coil. Preferentially, both devices would be
installed in a small expansion vessel that will allow also to fill the cryogenic vessel up to the
edge. Purification of the shielding water is needed to establish sufficient transparency for the 
detection of Cherenkov light; the radiopurity should be less than that of the construction
materials, stainless steel and insulation.

In the following section we will discuss some of the components in more detail and outline planned 
R\&D activities.

\subsection{Cryogenic tank options}

The generic design of the vessel is discussed within two limits requiring that the activity 
of the surrounding concrete (10~Bq/kg \thzza ) has to be suppressed by either a 
factor of 2.4 10$^{-9}$ or 2.4 10$^{-8}$. This will yield in the Ge energy spectrum between  2.00 and 
2.08~MeV a  background of $10^{-4}$ resp. $10^{-3}$ \ctsper , i.e. 
more than 3 resp. 2 orders of magnitude better than achieved in the Heidelberg-Moscow and \IGEX\ experiments 
($\sim$0.2 \ctsper ). 
\phantom{(see Table~\ref{ves-tab1})}
\begin{table}[h]
\begin{center}
\caption{\label{ves-tab1}
Linear absorption coefficients $\mu$ for 2.6~MeV \gam\ rays in various materials 
and the material's intrinsic \thzza\ activities assumed in the present discussion.
}
\vskip2truemm
\begin{tabular}{lcc}
\hline
\hline
material	&   $\mu$ [cm$^{-1}$]  &  activity [$\mu$Bq/kg \thzza] \\
\hline
LN		&  0.03115	&           \\
Water		&  0.0427       &      7000 \\
LAr		&  0.05		&           \\ 
Steel		&  0.299        &      7000 \\
Pb		&  0.484        &        30 \\
\hline
\hline
\end{tabular}
\end{center}
\end{table}

The vessel  dimensions are constraint by the available space in Hall A of the \LNGS\ to 
a diameter of 12~m and an effective height of 11~m. This space includes the volume needed
for the neutron moderator and the $\mu$ veto system. This leaves a height  of 
about 2.5~m on top of the vessel which is needed for the lock and cleanroom through which the 
Ge detectors can be installed in the vessel.

Cost and size of the vessel depend significantly on the choice of insulation. Most efficient would
be the use of superinsulation which at thermal losses of about 2~mW/m$^2$K needs marginal 
space, typically less than 20~cm. Its
implementation requires a special double-walled container which is able to maintain a vacuum of better
than $10^{-3}$~Pa between both shells. The more cost effective alternative is to take a normal
flat bottom tank as proposed for \geni\ \cite{kla99}. This device has a standard powder (perlite) 
insulation, which due to its much higher
thermal conductivity ($\sim$40~mW/Km) needs to be much thicker, typically 1~m.

Last not least, the vessel dimensions depend crucially on the choice of the shielding (and cooling) 
medium. 
If \lar\ would be used instead of \ln , the dimensions could shrink by the inverse
ratio of their densities, i.e. by a factor of 0.62, (see Tables~\ref{ves-tab1} and \ref{ves-shield-1}) 
and there would be, 
\begin{table}[h]
\begin{center}
\caption{\label{ves-shield-1} 
Thickness of various absorber materials needed to achieve the 
indicated background levels at indicated activities of steel and lead.
}
\vskip2.5truemm
\begin{tabular}{rlcc}
\hline
\hline
&	&   thickness [cm] for 		&   thickness [cm] for  \\
&	&  10$^{-3}$ [\ctsper ] 	& 10$^{-4}$ [\ctsper ]  \\
\hline
\hline
\multicolumn{4} {l} {\rule{0mm}{4mm}shield against concrete of 10 Bq \thzza/kg} \\	
\hline
1& LN 	&  563	&  637	\\
2&LAr 	&  350	&  396	\\
3&Fe~~~~~~~~~~~~~~~~\phantom{a} 	&  58.6	&  66.3	\\
4&Pb 	&  36.2	&  41.0	\\
\hline
\multicolumn{4} {l} {\rule{0mm}{4mm}shield against steel of 7 mBq \thzza /kg (3.34~cm = 1/e)} \\	
\hline
5&LN 	&  330	&  404	\\
6&LAr 	&  205	&  251	\\
7&Pb 	&  21.2	&  26.0	\\
\hline
\multicolumn{4} {l} {shield against steel and concrete (liquid - steel - Pb - concrete)} \\	
\hline
8&LN + Pb 	&  330+12.9	&  404+12.9	\\
9&LAr + Pb 	&  205+12.9	&  251+12.9	\\
\hline
\multicolumn{4} {l} {\rule{0mm}{4mm}shield against Pb of 30 $\mu$Bq \thzza/kg (2.07~cm = 1/e)} \\	
\hline
10&LN 	&  151	&  225	\\
11&LAr 	&  94	&  140	\\
\hline
\multicolumn{4} {l} {shield against Pb, steel, concrete (liquid - Pb - steel - concrete)} \\
\hline
12&LN + Pb	 	&  151+24.4	&  225+24.4	\\
13&LAr + Pb	&  94+24.4	&  140+24.4	\\
\hline
\hline
\end{tabular}
\end{center}
\end{table}
in fact,  
no problem to fit a LAr container in Hall A. Since we prefer to keep both options, 
we will discuss here optimized solutions which allow either the use of LN or LAr in an affordable 
way. For this purpose, we consider just those tank options which will provide with \ln\  a
background level of \dctsper\ and reach  \vctsper\ with a \lar\ fill only.

\subsubsection{Superinsulated vessel}
The realization of the original idea \cite{heu95} to use exclusively purified liquid 
nitrogen  as shielding material implies a LN thickness of 6.4~(5.6)~m
for arriving at the desired background level of $10^{-4} (10^{-3})$~\ctsper , 
see Table~\ref{ves-shield-1}. 
\begin{figure}[h]
\begin{center}
\includegraphics[width=10cm]{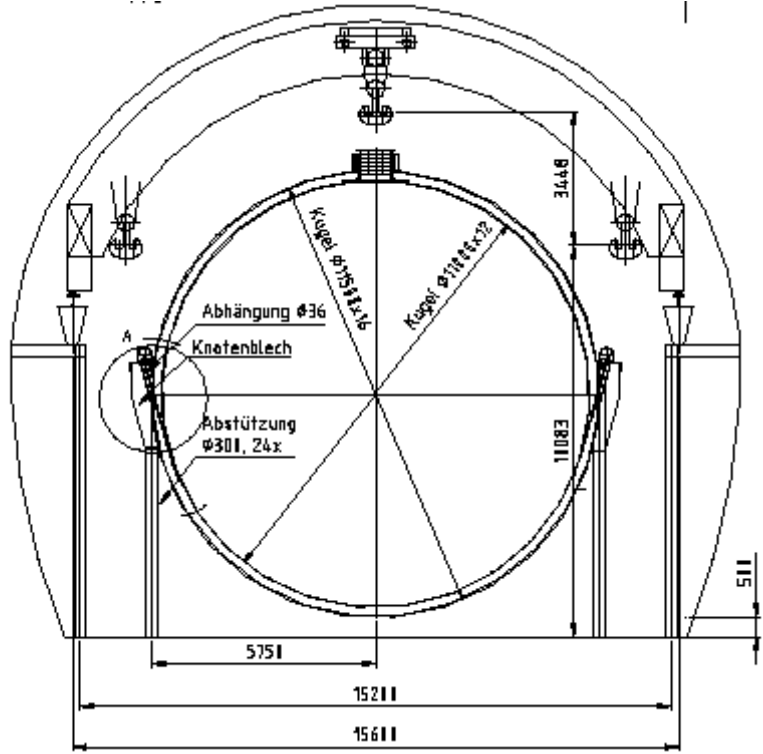}
\end{center}
\caption{\label{bab-ves} A vacuum insulated spherical LN vessel with an inner diameter of
11~m in Hall~A of \LNGS .
}		  
\end{figure}
A detailed study \cite{bab03} has shown
that it is possible to realize a vacuum insulated sphere with an inner diameter of 11~m 
(or a lying cylinder of 11~m diameter and 12~m length) that would fit into Hall A of \LNGS\  
(see Fig.~\ref{bab-ves}) and exhibit the structural stability needed to withstand an earth quake with 
an acceleration of 0.3~g. If the LN would be substituted by LAr, the dimensions would
shrink to a diameter of 8 (7)~m for a background of $10^{-4} (10^{-3})$~\ctsper . However, this solution
turned out to be prohibitively expensive, even disregarding the problem caused by the marginal space
available for the neutron shield.

The GEM project \cite{zde01} - a LN filled 
superinsulated copper sphere of 5~m diameter suspended in a water basin of 11~m diameter -
fits also in the category of superinsulated vessels.
It is not 
discussed here since the technical realization seems to be more difficult and expensive than a 
similar design described below.

\subsubsection{Flat bottom tank} 

\begin{figure}[h]
\begin{center}
\includegraphics[width=10cm]{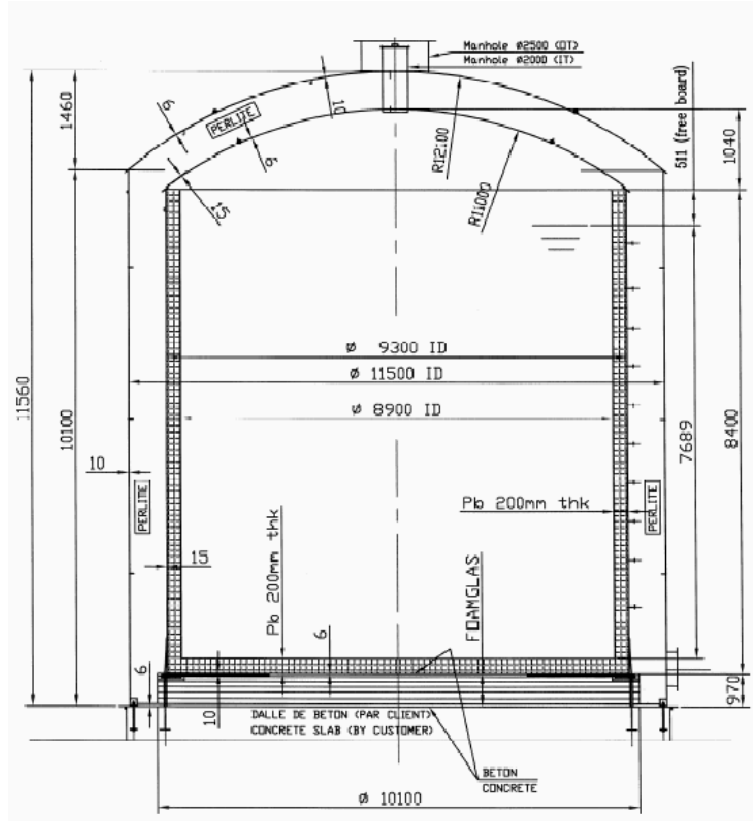}
\end{center}
\caption{\label{cmp-ves} 
Cross section of a commercial flat bottom vessel for 473~\cum\ LN or LAr 
with slight modifications to carry a non-standard internal Pb shield
(design by CMP Arles \cite{cmp04}). 
}		  
\end{figure}

The design of a flat bottom tank of the required shielding power with both LAr and LN fills
has to combine conventional and LN/LAr shields in order to fit into the available space.  
Table \ref{ves-tab1} lists the linear absorption coefficients for 2.6~MeV \gam - rays in LN and LAr
as well as in steel, lead and water. 
Accordingly, 6.44~cm of lead or 72.9~cm of water can substitute
a 1~m thick \ln\ layer. Water is an attractive substitute not only because it is cheap but
also since it can serve simultaneously as neutron absorber and Cherenkov medium for the detection
of muons.
In arranging the different materials, an additional constraint, however, is the
radiopurity of the involved materials.
Activities of 7000 resp. 30~$\mu$Bq \thzza/kg have been measured for  
the steel used in the LENS experiment and for commercially available lead, respectively, and
we adopt these values for the further discussion. In this context,
Fig.~\ref{cmp-ves} serves to illustrate typical features of a commercially available \cite{cmp04} flat 
bottom vessel; almost no modifications were made to account for the load by the non-standard internal 
Pb wall.  
The vessel's  base plate is bolted onto 1~m high concrete slabs supplied by the customer; 
(thus the vessel shown is $\sim$1~m too high to fit into Hall A). 
The $\sim$1~m thick insulation between the base plate and the inner vessel bottom consists of foam glass 
sandwiched between two 8~cm thick concrete layers. The remaining insulation is provided by a 1.1~m 
thick perlite layer. Table~\ref{ves-iso} shows measured \thzza\ activities for various insulation
materials. 
\begin{table}[h]
\begin{center}
\caption{\label{ves-iso}
Comparison of various isolation materials. The symbols A, $\lambda$, $\rho$, and $\xi$ denote
the \thzza(Th) radioactivity, thermal conductivity, density and compressive strength, respectively.
Class refers to German or British fire prevention classes.
}
\vskip2truemm
\begin{tabular}{@{}lccccc}
\hline
\hline
material	&      A    & $\lambda$  &  $\rho$    & $\xi$ & class \\
                &  [$\mu$B/kg]       &     [mW~/~Km ]        &  [kg/m$^3]$ &  [kg/mm$^2$] & \\
\hline 
Perlite \rule{0mm}{5mm}         &        3.5 $10^7$      &      45 - 70    &         &               &  A1    \\
Foam glass       &    	 1.0 $10^7$      &      40 - 60    &         &               &  A1   \\
Polystyrol (EPS)&         2100           &      30 - 40    & 15~-~30 &               &  B1  \\
Styrodur~C$^{\rm a}$ &                   &         40     & 25~-~45 & 200~-~700     &  B1   \\	
Elastopor$^{\rm b}$ &      d)        &        33.5     &         &    0.4        &  B3   \\	
Ecopir$^{\rm b}$&                        &                 &   33    &               &        \\
Ecophen$^{\rm c}$&                       &        18       & 60~-~160 &               &  0     \\        

\hline
\hline
\end{tabular}
\end{center}
\vskip1truemm
\hskip2truecm $^{\rm a}$ Polystyrol   \vskip-0.1truecm
\hskip2truecm $^{\rm b}$ polyisocyanurate/polyurethane rigid foam \vskip-0.1truecm
\hskip2truecm $^{\rm c}$ phenolic foam        \vskip-0.1truecm
\hskip2truecm $^{\rm d}$measurements in progress
\end{table}
Since foam glass exhibits the same activity  as the \LNGS\ concrete, 
there is no other choice but to install a lead shield together with the neutron 
absorber above the foam glass-concrete insulation, i.e. at the bottom of the cold volume. 
Table~\ref{ves-iso} shows in addition that at 30~Bq/kg the activity of perlite is even higher 
than that of the \LNGS\ concrete (10~Bq/kg). This rules out the use of this material. 
A substitute with even lower thermal conductivity is polystyrol.
Further candidates are  polyisocyanurate and phenolic rigid foams. Measurements are in 
progress to determine their activities. 
For the present purpose, the activity of the insulation is assumed to be equal or less than 
that of the steel.

Based on the above considerations and constraints, two natural options for the layout of a flat bottom 
cryogenic vessel emerge, and
Table~\ref{ves-shield-1} compiles the data for these options, assuming 
(A) installation of the lead  mostly outside the vessel or (B) completely within the cryofluid.
It is assumed  that the Ge detector assembly has a cylindrical shape of 50~cm diameter and
50~cm height, and that the insulation has a thickness of 1.1~m.
If not mentioned explicitly, the radiopurities of steel, lead, and water are those listed in 
Table~\ref{ves-tab1}.

\begin{figure}[h]
\begin{center}
\includegraphics[width=12cm]{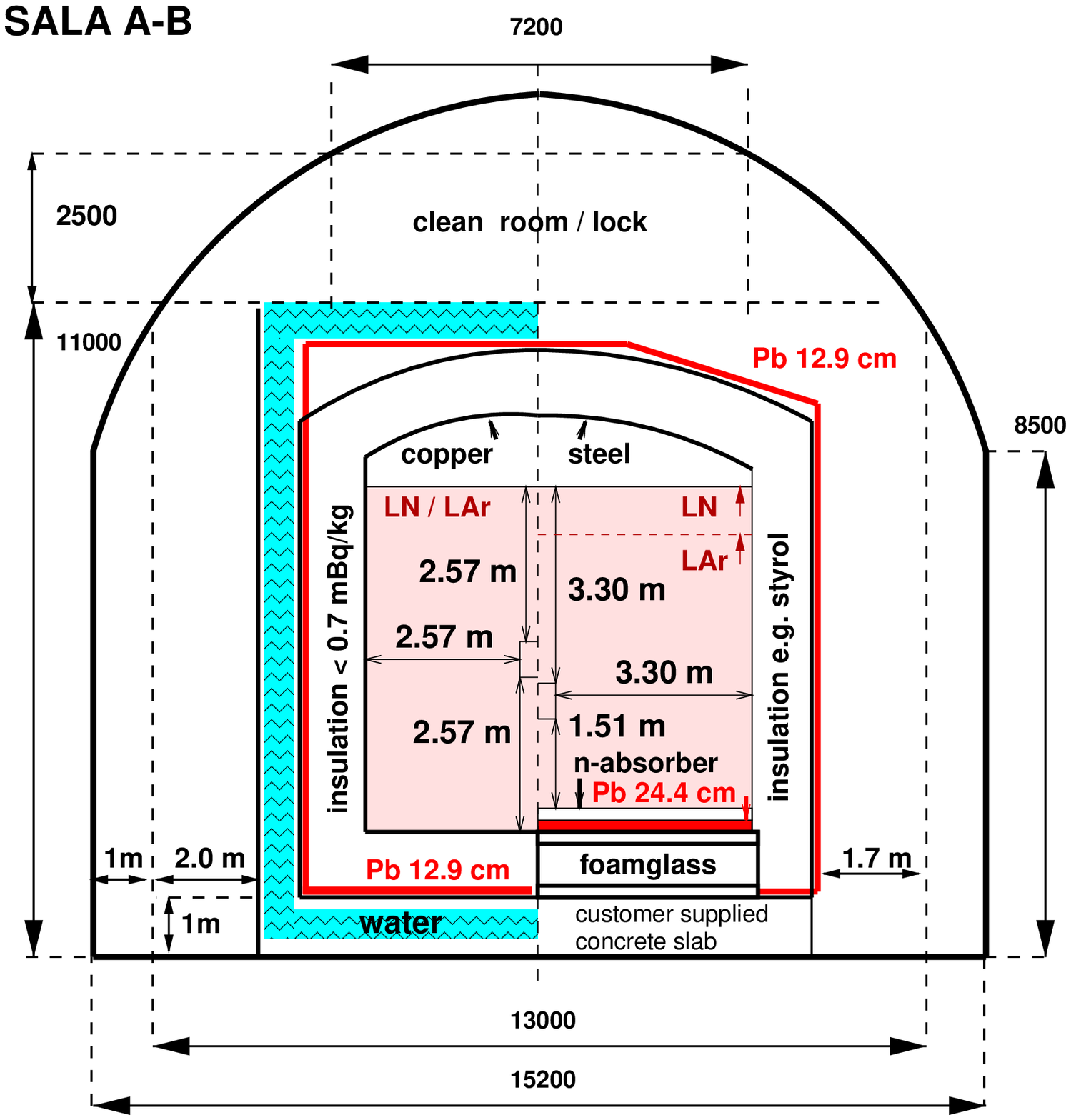}
\end{center}
\caption{\label{vopt-x} 
Possible layouts, options A (r.h.s.) and A' (l.h.s.), for a cryogenic vessel 
fitting into Hall A. 
The outer contour  shows the cross section of Hall A.
}		  
\end{figure}

\vskip0.3truecm
\noindent
{\bf Option A.} 
      The first alternative assumes standard commercial insulation at the bottom of the vessel, 
      i.e. foam glass and concrete, and a lateral insulation with an activity that is lower or 
      equal to that 
      of steel (7~mBq/kg \thzza). 
      The resulting vessel layout is shown at the right hand side  of 
      Fig.~\ref{vopt-x}. 
      Laterally and on top, the lead is placed outside the vessel.
      The minimum thickness of the cryofluid results from the constraint that it reduces the 
      background caused by the steel walls and the insulation to the desired level. 
      For a background level of \dctsper\ a thickness of 330~cm is needed for \ln . With \lar\ this 
      thickness is more than enough to reach a level of \vctsper\
      (see lines (8) and (9) in Table~\ref{ves-shield-1}). 
      The thickness of the additional lead needed to reduce  also the radiation from the 
      surrounding concrete to the desired level is 12.9~cm. 
      As discussed above, the high radioactivity of the foam glass  at the bottom enforces
      the lead shield to be placed here in the cold volume together with the neutron absorber, 
      e.g. a 30~cm thick polyethylene panel, placed on top of it. 
      The diameter and effective height of the vessel are about 10~m and 11~m, respectively. 
      These dimensions would allow one to replace part of the lateral lead shield by water.
      With LAr, the vessel would not need to be completely filled in order to yield a 
      background index of \vctsper .

\vskip0.3truecm
\noindent {\bf Option A'.} If materials of lower radioactivity were available, a much more 
          satisfactory layout would be possible as shown at the left hand side of 
          Fig.~\ref{vopt-x}. 
       The idea is to find a rigid foam of less than 700~$\mu$Bq/kg
       \thzza\ which would provide enough structural strength to allow for a thin inner
       vessel wall made of copper or low-activity stainless steel. Rigid foams of suitable 
       strength and thermal conductivity are indeed on the market - the still open question 
       is their radiopurity. Relative to option A the \ln\ thickness can  be reduced further   
       by 73~cm, the lead can be placed on the warm side of the insulation volume just in 
       front of the outer
       steel wall, and the remaining shielding power can be provided by a layer of water whose 
       thickness (40~cm) is equivalent to the 73~cm of \ln. Obviously this layer requires no extra
       space but a third wall.

\begin{figure}[h]
\begin{center}
\includegraphics[width=12cm]{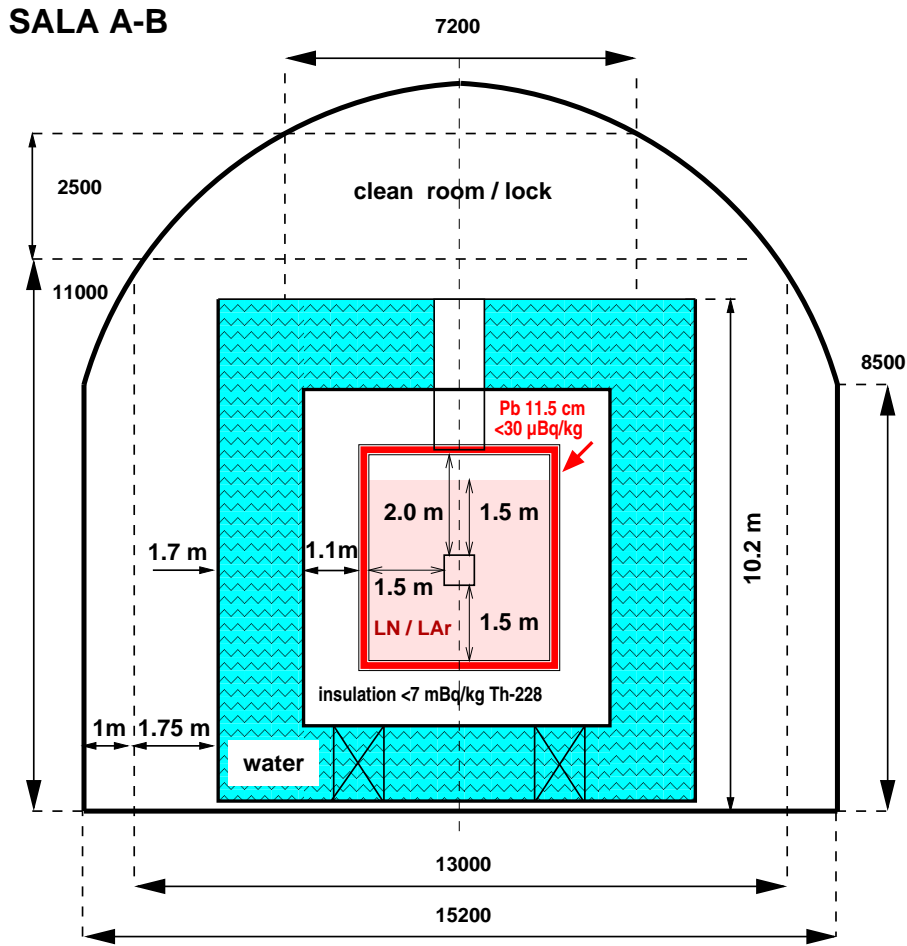}
\end{center}
\caption{\label{vopt-xs} 
Alternative layout for a cryogenic vessel fitting into Hall A:
the option B. 
The outer contour shows the cross section of Hall A.
}		  
\end{figure}

\vskip0.3truecm
\noindent
{\bf Option B.}
       This alternative leads to a most compact cryogenic vessel as shown Fig~\ref{vopt-xs} (see
       also Fig.~\ref{fig:generaldesign}).
  Starting point is the configuration of line (12) in Table~\ref{ves-shield-1} with the
  lead inside the cryofluid immediately in front of the inner wall.
  Studies of a specialized commercial company show the feasibility of this concept. The lead wall 
  can be stacked such that it is self-supporting and exhibits the required stability against earthquakes 
  \cite{jlg04}.   
  
  To shield the lead  of 30~$\mu$Bq/kg \thzza(Th) activity, a \ln\ thickness of 151~cm is needed for
  a background level of \dctsper .
  The thickness of additional lead required to reduce  also the radiation from the 
  surrounding concrete to the desired level is 24.4~cm. For obvious reasons, one
  would choose the lead thickness in the cold volume as small as possible, 11.5~cm,
  to shield the activity of the steel wall and the insulation. The remaining 12.9~cm could 
  be mounted outside the cold volume or, in view of the largely reduced overall dimensions, 
  substituted by a 171~cm thick water layer.
  The obvious savings in space and price might be even larger if the insulation thickness can be cut 
  roughly in half: with a total diameter of less than 5~m, the cryogenic vessel could be manufactured
  at a company and transported as a unit to \LNGS .
  The reduction of insulation thickness could be realized with evacuated powder insulation being 
  standard for tankers, or with superinsulation. In fact, a Russian company offers at a competitive 
  price a standard superinsulated vessel with an inner diameter of 3~m and suitable height between 
  4 and 6~m. 
  In case of  a superinsulated vessel made out of copper of $\leq 30~\mu$Bq/kg \thzza\, the lead 
  could even be placed outside the cold volume. 

\subsubsection{Conclusions}

Despite the many attractive features, superinsulated vessels show - with the exception discussed above - 
for 
the intended use no real advantage when compared to flat bottom tanks; they are not only more expensive 
but also not so flexible if instrumentation of the LAr is considered. 
For this purpose, an upright cylindrical flat bottom vessel with well distributed locks in the roof 
is close to the ideal solution. 

If one would discard the LN option in exclusive favor of LAr, a conventional flat bottom 
vessel of 10~m diameter, see line (2) of Table~\ref{ves-shield-1}, with little lead shielding at 
the bottom would be suitable. Part of the LAr 
could be even replaced by a water layer for neutron moderation and $\mu$ vetoing. When filled
with LN however, this vessel would yield only a background level of order 0.2~\ctsper , i.e. the level
of the Heidelberg-Moscow experiment.

If both the LN and LAr options are to be kept, several alternatives exist
depending on the radioactive purity of the insulation materials.
\begin{table}[h]
\begin{center}
\caption{\label{comp-opt}
Characteristics of three  cryogenic vessel layouts; for indicated options see also 
Figs.~\ref{vopt-x} and \ref{vopt-xs}.
}
\vskip2truemm
\begin{tabular}{llcccc}
\hline
\hline
&&\multicolumn{4} {c} {Option}     \\
\cline{3-6} 
\rule{0mm}{4mm}
Item            		& &     A & A' & B & \geni /\GEM\ \\ 
\hline 
\rule{0mm}{5mm}
\O$\times$H                    &[m$^2$]& $10\times 11$ & $9\times 11$ & $10\times 10$ & 
                                                                            $14\times 19/11\times11$\\ 
LN/LAr volume 			&[\cum ]& 210~/~178   &  141     &  34 & 1250/50 \\
Mass of Lead          		&[ton] & $\sim$550   &$\sim$500 &  98 &  0/0	\\
Volume of water 	     	&[\cum ]&    -         & $\sim$145         & $\sim$500 & 0/1000 
\rule[-2mm]{0mm}{1mm}
\\
\hline
\hline
\multicolumn{2} {@{}l} {\rule{0mm}{4mm}
Required radiopurities:} \\
\cline{1-2} 
inner wall    		        &&     7        &  0.7       & 7  &[mBq/kg \thzza]\\
insulation             		&&     7        &  0.7       & 7  \\
foam glass    	      		&&   10000      &  -         & -  \\
outer wall             		&&     7        &   7        & 7  \\
lead at bottom		      	&&   0.03       &  0.7       & 0.03 \\
lead elsewhere		      	&&     7        &  0.7       & 0.03 \\
water                           &&     -        &   7        & 7    \\ 
\hline
\hline
\end{tabular}
\end{center}
\end{table}
Table~\ref{comp-opt} gives a summary of the options discussed above as well
as the requested radiopurity of employed materials. 
Compared to \geni , all options do not only
need less space but exhibit a largely reduced volume of cryofluid. 
Options A' and B, in particular, integrate with a water shield also the neutron absorber in a 
space-saving way into the layout. 
In many respects, option B appears most attractive, 
also from the safety point of view. It uses the lowest amount of lead which is \lq cold\rq, 
however. 
Apart from a realistic costing which can be established only in close contact with industry, 
it is essential for the final decision to know if the close distance between lead wall and Ge 
array has any 
detrimental effect upon the neutron flux at the detectors. 

\LVD\ needs a passage for a tank carrier that is about 2.2~m wide
and 6.2~m long. This constraint will be taken into account by placing the LN/LAr vessel slightly
asymmetric into Hall A.

\subsection{Detector suspension} 

\begin{figure}[ht]
\begin{center}
\includegraphics[width=10cm]{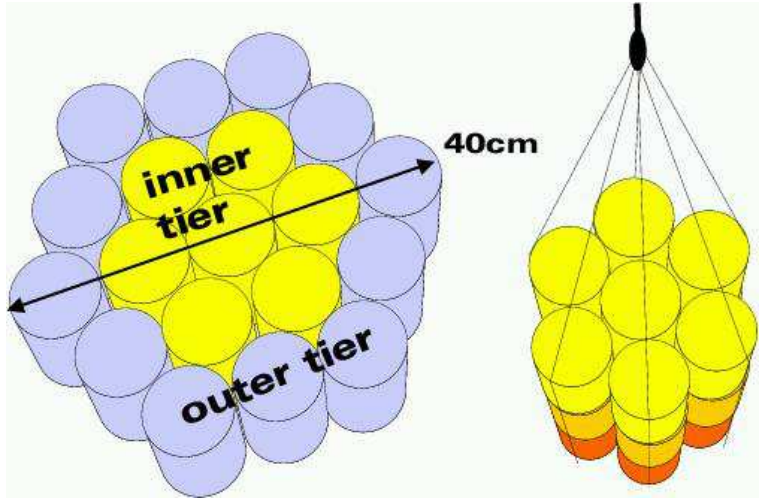}
  \caption{
  The detectors are arranged into a hexagonal structure.
  The inner tier contains 7, the outer 12 detectors.
  The picture on the left shows one complete layer with inner and
  outer tier. The one on the right depicts three layers of the
  inner tier with some suspension cables. 
}
  \label{susp1}
\end{center}
\end{figure}

The detectors will be integrated in assemblies allowing for
precise relative positioning and modular insertion.
The construction is to  minimize material while
using only low activity materials.
The complete system has to hold up to 50 detectors, each
with a mass of approximately 2kg. Not all detector crystals will
have the same dimensions. This will require special measures
to fix the relative positioning.
The electrical cables have to be guided in the system.
The left picture in Fig~\ref{susp1} shows a possible arrangement of detectors into
an inner and outer tier. The hexagonal arrangement minimizes the
gaps between detectors. The 7 detectors in a layer of the inner tier have
a total mass of about 14~kg. The 12 detectors in the outer tier
add 24~kg. Three complete layers would contain 112~kg of germanium.

There are several possibilities to divide the assembly into
modules that can be installed and exchanged separately.
Individual ``vertical strings'' are conceivable as well as
complete inner and outer tier assemblies that can be installed
independently.  
A cage like construction would guarantee a well defined  positioning of 
the individual diodes as well as an easy exchange; however, 
it is not clear if such a solution can be realized with the required 
minimum of material.

\subsection{Lock and cleanroom}

\begin{figure}[h]
\begin{center}
\includegraphics[width=8cm]{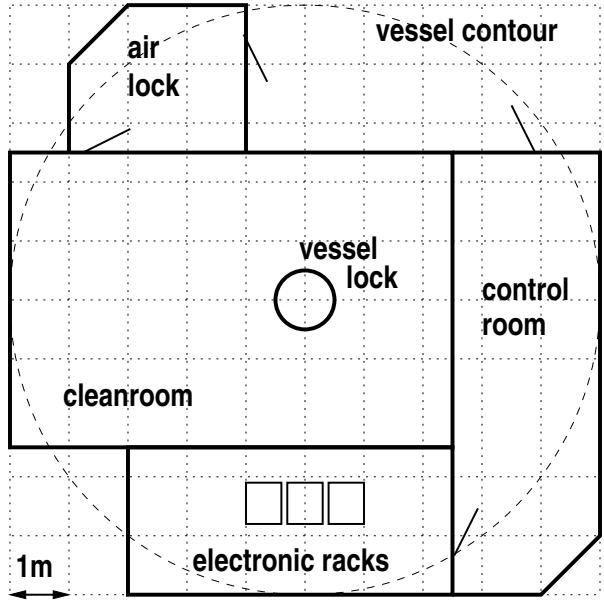}
\end{center}
\caption{\label{ontopof} 
A possible plan of the facilities on top of the vessel including 
vessel interlock, clean room, electronic hut and on-site control room.
}		  
\end{figure}

On top of the vessel a cleanroom at slight overpressure with an integrated lock 
provides the possibility to insert and withdraw detectors in a modular
way without contaminating the vessel's contents.
To keep
cable lengths short, the frontend electronics, power supplies and the PCs for
data acquisition and slow control are placed in a directly adjacent neighbored cabin.
Fig. \ref{ontopof} shows a possible floor plan that includes another cabin 
for the on-site control room.

The cleanroom is used for the final preparation of the detectors
and their integration into a modular detector suspension system.
The detector assemblies are transferred into the main lock
through a cleaning pre-lock where they are taken up by a rail system.
After the final cleaning procedure the assemblies can only be handled
through glove boxes integrated into the main lock. Cables and connectors
will be adjusted to that.

The main lock will be operated with a gas atmosphere that is as pure
as the vessel gas. 
In oder to prevent radon contamination the glove boxes will
be sealed when not in operation. The goal is to keep the lock as
clean as possible. However, during data taking  the hatch 
between lock and vessel will be UHV sealed to prevent any long term
contamination.

The main hatch will be 80~cm in diameter. The thermally insulated
flange will hold the mechanical suspension  system and the electrical 
feed-throughs for high voltage, power supply and signal cables.
An additional lead cover will provide shielding in the hatch region.

\subsection{Neutron and muon shield}
The cryogenic vessel is encased by a water layer with at least 40~cm thickness.
Its purpose is threefold: 
(i) to moderate and absorb neutrons, (ii) to attenuate the \gam\ flux, and (iii)
to serve as Cherenkov medium for the detection of muons crossing the lateral
wall of the experiment. The latter task implies the water to be purified 
(see subsection~\ref{sec:waterpurification}) to assure best transmission of the Cherenkov light.
The attenuation length of light should be larger than 10~m at $\sim$400~nm wavelength.
To maintain high transmittivity and radiopurity the water vessel has to be air tight. 

The Cherenkov light is detected by photomultiplier tubes (PMTs), for example
the 8" Hamamatsu R1408 PMT that in case of Super-Kamiokande \cite{suk03} is covered 
with a wave length shifter plate of 60~cm$^2$ area.
Another effective means to increase the number of detectable photoelectrons
is to increase the reflectivity of the walls of the water tank. It has been shown, 
see e.g. \cite{sho99}, that a lining with Tyvek, a fibrous polyolefin manufactured 
by Dupont, provides a diffusively reflective surface with greater than 90\% reflectivity  
at the relevant wave lengths around 350~nm. 

Since the detailed geometry of the water tank is not yet determined it is not possible to
discuss the arrangement of the PMTs in detail. Monte Carlo simulations will be used
to study the muon detection efficiency as a function of number and arrangement of the 
PMTs. A first estimate shows that about 150 PMTs will be necessary to detect muons with 
sufficient efficiency. 
These PMTs could be encapsulated together with their remote controlled HV converters, amplifiers and 
discriminators in watertight acrylic spheres of 12 inch diameter. 
Strings of such optical modules would be
hung down from the top of the water vessel. The PMTs will be working in the single 
photo-electron regime, and their dark current will imply a counting rate of 10~kHz. Hence
a 4-fold 10~ns wide coincidence between neighbored optical modules will be used to identify the 
signal of an incident muon. The calibration system will consist of a pulsed UV laser whose output
is coupled by optical fibers to the individual PMTs.    

Of primary importance is the coverage of the top of the vessel. For performance
and logistic reasons, it might here be desirable to choose, instead of water,   two layers of 
scintillator plates which are operated in coincidence.

\subsection{Electronic readout}

\begin{figure}[h]
\begin{center}
\includegraphics[width=13cm]{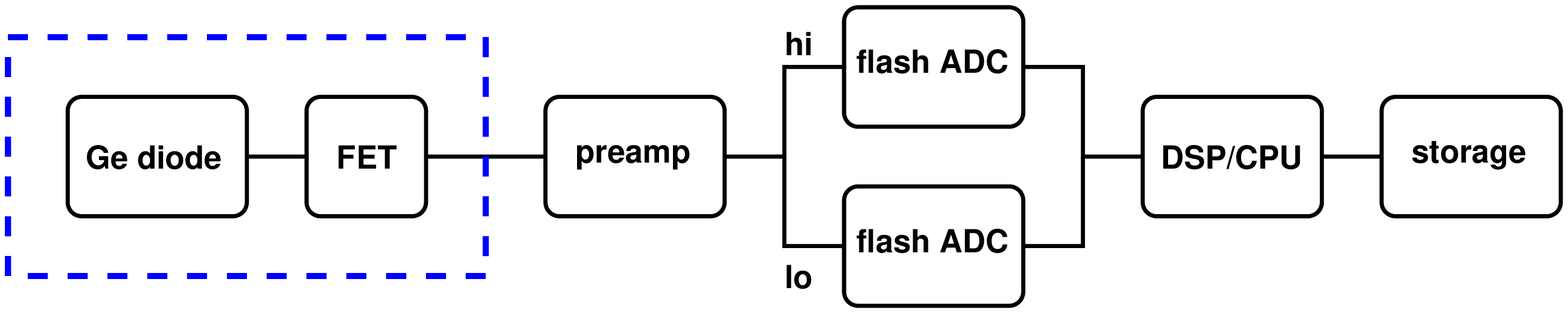}
\end{center}
\caption{\label{elecread} 
Schematics of the electronic readout chain.
}		  
\end{figure}

Figure~\ref{elecread} shows a schematic of the readout chain for one Ge diode.
For optimum noise and speed the input FET of the charge integrating preamplifier 
is placed close to the detector within the cryofluid. The pre-amplified signal
is sampled by a flash ADC (FADC); two such units of high and low gain may be needed
if the dynamic range of one unit is not sufficient. Further signal processing 
including filtering will be done fully digital by digital signal processors (DSPs),
field programmable gate arrays (FPGAs) or standard PCs. 
Algorithms for the real time digital synthesis of pulse shapes 
have been established \cite{jor94,jor03} for deriving the optimum information on energy and time 
from a given signal.

Besides the readily available and well proven commercial products for FET and preamplifier,
the developments for \gam\ ray tracking detectors like \AGATA\,\cite{aga} have resulted in 
frontend electronics optimized for pulse shape analysis with segmented Ge detectors. 
Besides discrete hybrid solutions also fully integrated ASIC designs are pursued \cite{del03}. 
Typical specifications include a noise 
level of 1~keV FWHM, a bandwidth of 20 to 30~MHz, a dynamic range of 1000, and a FET power of less 
than 25~mW which makes these frontends extremely interesting also for the present application.

For the pulse shape sampling and digital signal processing units, several options are available.
The commercial fully digital spectroscopy modules of XIA's DGF4C series \cite{xia} have been 
chosen as a reference by the \Majorana\ collaboration \cite{maj03}. The CAMAC based DGF-4C as well
as the DGF Pixie-4 module in the CompactPCI/PXI standard  provide digital spectrometry and waveform 
acquisition for four input signals per module with the possibility to combine several modules 
into a larger system. In the Pixie-4 the signals are digitized in a 14 bit FADC at 75~MHz.
Triggering, filtering and time-stamping of the data stream is done in real time in a field 
programmable gate array (FPGA), and the resulting data can be read out by a computer at up to
109~MB/s. An alternative more cost-effective solution is represented by the GRT4 VME pulse
processing card which has been developed for the determination of position, energy and time
of an \gam\ event in a segmented Ge detector \cite{laz03}. Each of its four channels has a 40~MHz 
low pass filter including an optional differentiation stage followed by a 14~bit 80~MHz FADC.
The VHDL hardware language is used to implement the desired algorithms in the FPGA. An in-beam
test with a 33\% efficient Ge detector has shown that the energy resolution is the same as
with conventional analog electronics.  A third even more cost-effective solution \cite{kih03} uses 
the 8-channel SIS3300 VME FADC module with 100~MHz sampling rate at 12 bit precision.   
The module is self-triggering and has two memory banks which allow data transfers at 7~MB/s.
A low noise amplifier in front of the FADC is used to adjust gain and dc offset.
All digital signal processing including pulse shape analysis, integration and pole/zero 
cancellation was done by a VME CPU with an Intel p3/850~MHz processor running a Linux operating
system. This system showed a similar energy resolution as an analog system at a  maximum count 
rate of 4000 events/s and has been installed in the \geni\ test facility at \LNGS\,\cite{kla03a}. 

In conclusion, various attractive readout systems are existing or are becoming available. The
final choice will be made after extensive tests and a detailed cost/performance evaluation.  

\subsection{Data acquisition and Slow Control}

The data acquisition system will gather the data of the Ge diodes and of the muon counters, and
store them with a time stamp on tape. 
Neither channel count nor acquisition rate will represent a problem - even the muon count rate is only
several 10$^{-4}$ events per (m$^2$s). The major requirement will be rather the highest possible 
operational availability and stability. 

The slow control system handles traditionally the non-time critical tasks. It transfers
the whole detector system from the safe stand-by mode into the running state (or vice versa)
by activating the low voltage power supplies and setting the high voltages for the photomultipliers
and Ge-diodes. Concurrently it is reading at typically a few Hertz the various system parameters, 
and stores any changes of their values with a time stamp in a data base for later retrieval. It also 
provides a graphical user interface for a breakdown of the status of the various subcomponents, as
well as for loading experiment configuration files.
Typical system parameters include 
\begin{itemize}
\item from the vessel: pressure, temperature, gauge of cryogenic fluid and water,
\vskip-10truemm
\item from the detectors: leakage currents and base currents of the photomultipier tubes, 
\vskip-10truemm
\item from the electronics: power supply voltages and currents, status of crates, temperatures, 
\vskip-10truemm
\item from the cleanroom: air pressure, oxygen level, radon level, particle concentration, 
\vskip-10truemm
\item from the environment: barometric pressure, humidity, temperature, oxygen level, status
                                of smoke detectors.
\end{itemize}				
If safety relevant alarm conditions are detected, they are immediately communicated to the 
programmable logic controller (PLC) which forwards them to the \LNGS\ general safety monitoring 
system. For most safety critical parameters, however, the detour via slow control is avoided and the
monitoring hardware will send its  alarm signals directly to the PLC.

The full computerization of the data acquisition and slow control system will allow 
the  sharing of acquired data among all collaboration members practically in real-time, as well
as the reliable remote monitoring and control of the whole experiment.

\subsection{Liquid gas purification}

\newcommand{\BX}{\mbox{{\sc Borexino}}}
\newcommand{\LENS}{{\sc Lens}}
\newcommand{\STRAW}{{\sc Straw}}
\newcommand{\Gallex}{{\sc Gallex}}
\newcommand{\GEMPI}{Ge{\sc mpi}}
\newcommand{\Rn}{$^{222}$Rn}
\newcommand{\Ra}{$^{226}$Ra}
\newcommand{\Ar}{$^{39}$Ar}
\newcommand{\Kr}{$^{85}$Kr}
\newcommand{\Xe}{$^{133}$Xe}
\newcommand{\Yb}{$^{176}$Yb}
\newcommand{\Fe}{$^{55}$Fe}
\newcommand{\Th}{$^{231}$Th}
\newcommand{\Ul}{$^{235}$U}
\newcommand{\Uh}{$^{238}$U}
\newcommand{\Be}{$^7$Be}
\newcommand{\YO}{Yb$_2$O$_3$}
\newcommand{\C}{$^\circ$C}
\newcommand{\al}{$\alpha$}
\newcommand{\be}{$\beta$}
\newcommand{\ga}{$\gamma$}
\newcommand{\dop}{\mathrm{d}}  
\newcommand{\spa}{\ \;}        
\newcommand{\ta}  {\rule[-1,5mm]{0mm}{7,5mm}}
\newcommand{\tag} {\rule[-3mm]{0mm}{9mm}}
\newcommand{\up}  {\rule{0mm}{6mm}}
\newcommand{\down}{\rule[-3mm]{0mm}{3mm}}

Liquid nitrogen (\ln) and liquid argon (\lar)  are produced by air separation and thus contain traces of 
atmospheric noble gases.  The most abundant radioactive noble gas nuclides in the atmosphere are 
\Ar, \Kr\ and \Rn\ (see Table \ref{atmosphere}).  \Ar\ and \Kr\ have Q-values below 700~keV and cannot 
contribute to the background in the $0\nu\beta\beta$-region. 
\begin{table}[h]
\caption{\label{atmosphere}
The most abundant radioactive noble gas nuclides in the atmosphere 
(all gas volumes are given at STP).
}
\begin{center}
\vskip2truemm
\begin{tabular}{|c|c|c|c|}\hline
\tag & \Ar & \Kr & \Rn\\ \hline

\up     Activity & 1.8 Bq/m$^3$ Ar & 1.2~MBq/m$^3$ Kr &\\
\down concentration & 1.7 Bq/m$^3$ air & 1.4 Bq/m$^3$ air & 5 Bq/m$^3$ air\\
\hline
\tag Reference & \cite{loo83} & \cite{bfs01} & \\ \hline
\end{tabular}
\end{center}
\end{table}
 
The \Rn-decay chain is more complex.  
In $^{214}$Bi and $^{210}$Tl decays, $\gamma$ rays of more than 2~MeV energy are emitted.
Monte-Carlo-simulations have shown \cite{bau03} that 
a \Rn\ activity of 0.5~$\mu$Bq/m$^3$ of nitrogen (STP) leads to a count rate of 
$6\cdot 10^{-5}$~\ctsper\ in the energy window between 2~MeV and 2.08~MeV.  
Consequently nitrogen of this \Rn-purity is sufficiently pure for the experiment.

Similar purity levels are required for liquid argon. When the scintillation of argon is exploited
it might become necessary to remove \Kr\ from the argon in order to reduce the overall count rate.  
The \Kr\ activity should not exceed the \Ar\ volumetric activity of 1.8~Bq/m$^3$ (STP) .  
Therefore the volumetric krypton concentration in the argon has to be 1~ppm or lower.  

\subsubsection{Noble gas adsorption}

Ultra pure gases are produced by adsorption of the impurities on dedicated adsorbers.  
For low partial pressures $p$ the number of mols $n$ of a gas that are adsorbed per mass of 
adsorber is given by Henrys law: $n = H \cdot p$~; $H$ is the Henry constant.  
Since activated carbon usually has a wide pore size distribution, it is suited to trap different 
kind of impurities independent of their molecular size.  
\cite{mau00} gives the following empirical equation for the single component adsorption of 
un-polar gases on activated carbon:

\begin{equation}
H\Bigg[\frac{\mathrm{mol}}{\mathrm{kg\!\cdot\!Pa}}\Bigg]=
\exp\Bigg\{\Bigg(-0,05+\frac{81}{T [\mathrm{K}]} \Bigg)\!\cdot\!\frac{T_C [\mathrm{K}]}
{\sqrt{p_C [\mathrm{bar}]}}-17,5\Bigg\} \label{tCpC}
\end{equation}

$T$ is the temperature and $T_C$, $p_C$ are the critical temperature and pressure of the adsorbed gas.  
Table \ref{kritisch} shows the critical parameters for the gases of interest and their calculated 
Henry constants for the adsorption on activated carbon at different temperatures.
\begin{table}[h]
\caption{\label{kritisch}
The critical pressures and temperatures for different gases and the calculated Henry constants for 
the adsorption on activated carbon.
}
\begin{center}
\vskip2truemm
\begin{tabular}{|c|c|c|c|c|c|c|c|}\hline
\up Gas & $T_C$ & $p_C$ & $T_C$/$\sqrt{p_C}$ & \multicolumn{3}{|c|}{$H$ [mol/(kg$\cdot$Pa)]} & Ref.\\ \cline{5-7}
\down & [K] & [bar] & $\Big[$K/$\sqrt{\mathrm{bar}}\Big]$ &  288~K & 173~K & 77~K & \\ \hline

\up     Ar & 150,7 & 48,6 & 21,6 & $4 \cdot 10^{-6}$ & $ 2 \cdot 10^{-4}$ & 63 & \cite{atk96}\\
  N$_2$ & 126,3 & 34,0 & 21,6 & $4 \cdot 10^{-6}$ & $ 2 \cdot 10^{-4}$ & 63 &\cite{atk96}\\
   
      Kr & 209,4 & 55,0 & 28,2 & $2 \cdot 10^{-5}$ & $3 \cdot 10^{-3}$ & $5 \cdot 10^4$ & \cite{atk96}\\

\down   Rn & 377,0 & 62,8 & 47,6 & $2 \cdot 10^{-3}$ & 10 & $1 \cdot 10^{13}$ & \cite{fle03}\\ \hline

\end{tabular}
\end{center}
\end{table}

Being by far the heaviest gas of highest polarizability, radon  can already be adsorbed efficiently at 
rather high temperatures (173~K).  At liquid nitrogen temperature the Henry constant for radon 
becomes huge, so the radon is completely transfered from the gas phase to the adsorbed phase.  
Therefore no radon should remain in liquid nitrogen after running over an activated carbon 
column at 77~K.  In practice re-contamination due to \Rn\ emanation of the activated carbon limits 
the obtainable purity.  The \BX\ collaboration uses the described technique to produce high purity 
nitrogen.  The emanation problem could be solved by using a \Ra-free synthetic carbon 
(\Rn\ emanation rate (0.3 $\pm$ 0.1)~mBq/kg \cite{heu95}).  In this case no radon was detected in the  
purified nitrogen.  The obtained upper limit is 0.3~$\mu$Bq/m$^3$ (STP) \cite{mpi03}.\\

The elimination of krypton is more difficult.  
Due to the similar adsorption properties of krypton and nitrogen/argon 
(see Table \ref{kritisch}) a single component adsorption model fails to predict the equilibrium.  
The situation can be improved substantially if the adsorption happens in the gas phase at temperatures 
slightly above the boiling point of the gas to be purified. Further improvement is possible if carbon 
adsorbers with optimized pore size distributions are used.  
Doing so it is possible to achieve Henry constants for krypton of the order of 
1~mol/(kg$\cdot$Pa) at 95~K in the binary system krypton/nitrogen,
and to purify more than 500~m$^3$ of nitrogen (STP) from krypton with only 1~kg of adsorber.  

For argon the ratio of $T_C$ and $\sqrt{p_C}$ is the same as for nitrogen.  
Therefore the purification of argon from radon and krypton can be performed in the same way as the 
nitrogen purification.  
 If the argon is to be instrumented, residual oxygen
         is of concern as it can quench the scintillation efficiency.
The {\sc Icarus} collaboration has successfully operated a commercially available oxygen 
purification unit with liquid argon \cite{cen93}.  
 However, it is known that this material emanates radon. 
         If such a device is  implemented, it
	         has to be located  up-stream of the carbon adsorber to remove
		         the emanated radon. In any case, the radon levels of the
			         various components will be assayed prior to the
				         finalization of the design.

The purification of nitrogen from argon is impossible by adsorption, because their adsorption 
properties are almost identical (see Table \ref{kritisch}).  A separation can only be achieved 
by rectification as it is done during the production of nitrogen in air separation plants.  
The \BX\ collaboration has investigated high purity nitrogen from different suppliers \cite{zuz04}.  
The best quality, see Table~\ref{westfalen}, has a volumetric argon concentration of 0.5~ppb,
\begin{table}[h]
\caption{\label{westfalen}
Contaminations in commercially available liquid nitrogen.
}
\begin{center}
\begin{tabular}{lccc}
\hline
\hline
Company	&        Ar [ppm]  &  Kr [ppt] & $^{222}$Rn [$\mu$Bq/\cum ] \\
\hline
Air Liquide (4.0) & 10     & 40   & $\sim$50 \\
Linde (7.0)       & 0.06   & 0.3  & ? \\
SOL (6.0)         & 0.005  & 0.04 & ? \\
Westfalen AG (6.0)& 0.0005 & 0.06 & ? \\
\hline
\hline
\end{tabular}
\end{center}
\end{table}
and is  pure enough for the experiment.  Moreover the volumetric krypton concentration 
was 0.06~ppt.  
Therefore, it might not be necessary to build purification plants for krypton and argon.  
Radon, however, has to be removed before the detector is filled because it is emanated from most 
surfaces so a re-contamination can not be excluded. 
Since both \ln\ and \lar\ are produced by the same process,
rectification, one might expect  that the krypton contamination of commercially available Ar~6.0 is 
less than 1~ppm, a value which should be verified by measurement, however.

\subsubsection{Monitoring of liquid gas purity}

Techniques to determine the amount of radon, krypton and argon in nitrogen 
have been developed for the \BX\ experiment.  
\Rn\ in nitrogen can be measured at the sub-$\mu$Bq-level with proportional 
counters \cite{heu95}.  Argon and krypton concentrations are determined with a 
specially tuned rare gas mass spectrometer.  
The detection limits for 1~ccm nitrogen samples are $\sim$1~ppb for argon and 
$\sim$0.1~ppt for krypton \cite{zuz04}.  

The detection of radon in argon can be done in the same way as described in \cite{heu95}, 
replacing nitrogen by argon.  The radon from several 100 m$^3$ of argon will be trapped 
on an activated carbon column cooled at liquid argon temperature.  
After removing most of the argon the radon is filled into the proportional counter.  
To determine the krypton concentration in argon an extension of the technique applied for 
the determination of krypton in nitrogen is required, since mass spectrometers cannot 
handle macroscopic amounts of noble gases. It has been shown \cite{sim03} that
gas chromatography allows one to remove the largest fraction of argon, while at the same time
the traces of krypton can be separated  without loss \cite{sim03}.  
The extracted krypton can then be measured with a standard mass spectrometer.
 
\subsection{Water purification}
\label{sec:waterpurification}

The water shield will serve also as Cherenkov medium for the detection of incident muons.
Purification of the water is necessary in order to keep the light transmissivity as high as
possible. Origins of turbidity are dust and metal ions as well as biological activity from
bacteria, for example. Contaminations from radioactive materials like Rn, Ra and Th should
be less than the steel activity of 7~mBq/kg. 
 
Water purification systems are essential for water Cherenkov detectors like Super-Kamiokande
\cite{suk03}, Auger \cite{aug97} or the \BX\ counting test facility (CTF) \cite{ali98} 
where the design goal for the radiopurity of the water shield is $10^{-6}$~Bq/kg.
The radon content of the \LNGS\ water is 10$^4$~Bq/\cum .

The purification systems consist in general of several components including $\mu$m absolute 
filters, ion exchangers for the removal of metal ions, ultra violet sterilizers to kill bacteria, 
and ultra filters to remove nanometer sized particles \cite{aug97,ali98,suk03}.
The purity of the water is monitored by measuring in particular its resistivity which reaches
after the ultra filter the chemical limit of 18~M$\Omega \cdot$cm \cite{suk03}. 
Radon may be removed either by a vacuum de-gasifier \cite{suk03} or by a counter-flow 
of nitrogen like in the \BX\ CTF \cite{ali98}.

The capacity of the \BX\ CTF purification system is designed for 1000 tons of water.
In order to save cost and space, we propose to
use this system for producing the $\sim$500 tons of water needed for the planned experiment.
In view of the sealed and rust-protected water container, a small and dedicated continuous
re-circulation system with a sub-micron filter and an ultra violet sterilizer will be sufficient to
maintain the water quality.
 
\subsection{Liquid gas storage}

Storage tanks are needed to provide timely the \ln\ (or \lar ) lost by evaporation.
This loss will range - dependent on vessel volume and type of insulation - from 0.1 to 1~\cum\ per day.
Hermetic storage of freshly delivered liquid gas  over a few weeks is also a cheap yet efficient
method to reduce the content of radon by profiting from its half-life of less than 4 days.

Close to Hall A two storage tanks are located with a capacity of 5~\cum\ each. So far, they have been 
used by the GNO experiment. These tanks might be suitable for the planned experiment and we propose
to allocate them to it. However, their compatibility with the required low Ar, Kr and Rn concentrations
has to be checked.


\newpage

\section{Procurement of enriched \geenr\ detectors}

\subsection{Modification of existing \geenr\ detectors}

\begin{figure}[h]
\begin{center}
\includegraphics[width=12cm]{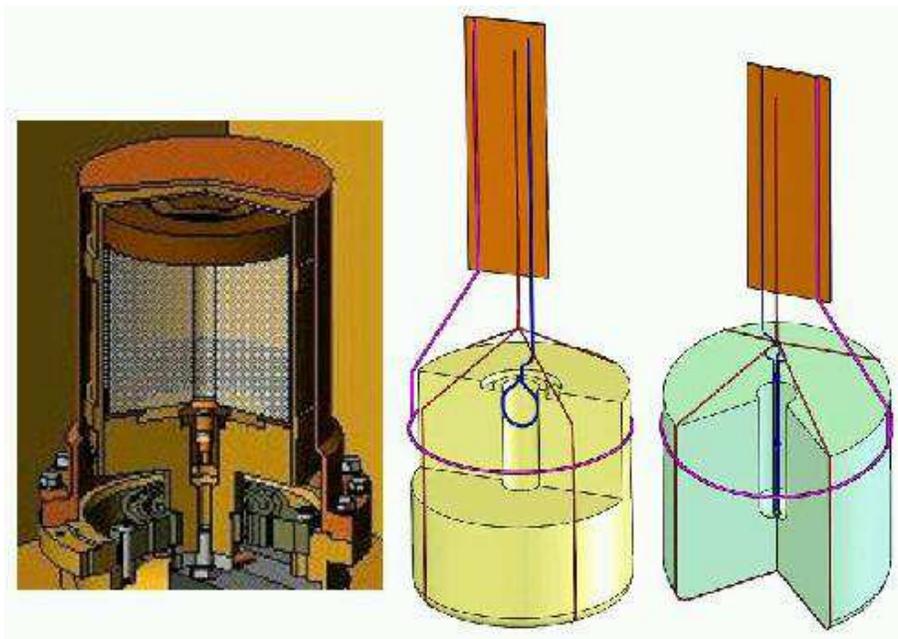}
\end{center}
\caption{\label{contacts}
The conventional way of a Ge-diode mount (left) and proposed suspension
and contacts for \IGEX\ type (middle) and HdM-type (right)
enriched \geenr\ diodes.
}
\end{figure}

There are available five, respectively three, enriched \geenr\ detectors from the 
Heidelberg-Moscow (HdM) and 
\IGEX\ collaboration with a mass of about 2~kg each. All these detectors are
of p-type and manufactured in a similar way. At present, these detectors are
still housed in cryostats stored underground either at \LNGS , Canfranc or Baksan.
For the use in the proposed experiment, it is necessary to take these detectors out of 
their cryostats (see Fig.~\ref{contacts}, left side)  and to remove the old contacts. 
After a cleaning procedure the detectors
need to  be equipped with new contacts. 
Two possible solutions are depicted in the middle and right part of Fig.~\ref{contacts}.
For the \IGEX\ type diodes, a steel spring wire of 0.5~mm diameter serves  
as inner signal contact. 
The suspension strings as well as the HV contact  are fixed by the same material which - by use of a 
tiny spanner nut - is tightly laid around the outer surface of the Ge cylinder. 
As to the inner contact of the HdM-type diodes, three legs of the steel spring wire are 
welded together and bound on top in such a way that they form a spring;
attached at the merging point of the suspension strings, the spring load presses the tip 
against the core bottom. 
This contact could also be installed upside down. 
Another alternative would be a hollow stem  made of electro-formed Cu with a spring coil on top. 
At the crystal interface the contacts will be gold plated.  However, the optimal low-mass contact 
solution would be to bond the wires directly to the crystal surface, and tests are in progress. 

The suspension strings consist of 0.5 mm Dyneema fiber. 
While the figure shows four legs, three should be sufficient. 
Both contacts and the central suspension  are integrated into a Kapton strip to 
assure fixed spacing (capacity) between the wires. 

The approximate weight of the materials close to the crystal hardly exceeds 2.6 g 
(spring wire $\sim$0.4 g, fiber $\sim$0.2 g, screw system close to 2 g). 
This compares with a mass of 3.5 kg for the HdM detectors which is mainly due to  the copper 
envelope and cap. 
The total surface area of the described solution is about 200 times smaller than in the 
conventional case. 
All materials to be used have been checked already by Ge screening at Heidelberg to be radiopure 
below a level of a few 10 mBq/kg. Measurements with higher sensitivity are planned. 
Mechanical stability tests of these materials at liquid nitrogen temperature have been  
performed with satisfying results.

Any effort will be made that all modifications are done underground in order to avoid a
new contamination of the crystals by cosmogenic radioactivity.

\subsection{Fabrication of enriched \geenr\ detectors}

The fabrication of intrinsic Ge detectors is usually the realm of commercial companies.
With respect to enriched detectors, however, several problems have to be overcome.
One problem is the procurement of enriched material which - at affordable cost - can only
be supplied by Russian companies; another problem is the cosmogenic radioactive contamination
of the \geenr\ material if it is kept at sea level and not stored immediately 
underground (see chapter \lq Background simulations\rq ).

\subsubsection{Procurement of enriched \geenr\ }

Information from the Ministry of Atomic Energy of Russia obtained by the end of 2003 and in 
January 2004
indicate that enriched Ge can be acquired from the Svetlana plant which is a special
department of the Zelenogorsk electro-chemical plant (ECP). Together with the
Kurchatov institute ECP has already supplied the 
enriched material for the Heidelberg-Moscow and \IGEX\ experiments. 
The separation is done in centrifuges for stable isotopes.
The necessary natural Ge has to be supplied by the customer 
(14~kg for 1~kg of 85\% enriched \geenr ). The preferred form of natural Ge is metal
granulate but metal bars are also accepted. Its purity should be $>$99\%. 
Typical and easily available qualities are 4N or 5N.
The enriched Ge will be delivered as GeO$_2$; at slightly higher cost, however, delivery as metal is
also possible. For each batch the contaminations will be specified. 
The Ge waste material can be returned on customer's request either as GeO$_2$ or Ge metal.

The minimum batch is  30 kg of enriched \geenr . At the beginning, however, a test run
yielding  about 2 to 4 kg enriched \geenr\ for material validation is possible. The maximum
production capacity is 200~kg per year, and the enrichment process will run without any
interruption. The price is 45 to 50 US\$ per gram 85\% enriched \geenr\ plus the cost
of the natural germanium.

Research at the Institute of Molecular Physics of the RCC Kurchatov is pursuing an alternative
separation process based on GeCl$_4$. 
Thus expensive
chemistry steps for the conversion of germanium into GeF$_4$  could be reduced resulting in
a presumably lower price for the enriched material (see section~\ref{sec:gecl4}).
 
\subsubsection{Processing of \geenr O$_2$ to intrinsic \geenr\ crystals}

In Europe, we know of only one  big  commercial company which does both the reduction 
of GeO$_2$ to metal, the subsequent zone refining to the level of a few 10$^{10}$ impurities 
per cm$^3$ required for the production of intrinsic Ge diodes, as well as the crystal pulling.
Different from the past, the reduction is no longer done in a wet chemistry environment but in a 
hot H$_2$ atmosphere which leads us to expect that the losses of enriched material become negligible in this step. 
The processing of
Ge material is carried out on large scale and it has to be explored how the separate processing of relatively
small amounts of enriched \geenr\ material would fit into the production flow. First encouraging
contacts have been made.    

\subsubsection{Detector fabrication}
  
There is no evidence that fabrication of intrinsic Ge detectors is made more difficult if isotopically 
shifted material is used. The enriched \geenr\ detectors of the Heidelberg-Moscow and IGEX
experiments, more than 10 detectors with an active mass of about 20~kg, have been manufactured by 
commercial 
companies and operated subsequently for more than 80~kg$\cdot$years.

Natural Ge (p-type) detectors which have proven to operate in \ln\ or \lar\ have been supplied by three
companies so far, and we see no reason that this success is due to any proprietary technique. 
\phantom{Table~\ref{time} shows }
\begin{table}[h]
\begin{center}
\caption{\label{time} 
Estimated time which has to be allocated to various detector processing steps.
}
\item[] \begin{tabular}{@{}ll}
\hline
\hline
Step &  time needed  \\
\hline
polyzone refining          & 2 days                        \\
mono-zone refinement incl. quality control meas. & 7-8 days \\
crystal pulling incl. last zone refining &  2 days \\
\hline
production of standard 2kg Ge diode      & approx. 2 weeks \\
\hline
\hline
\end{tabular}
\end{center}
\end{table}

The major constraint for the fabrication of \geenr\ detectors for 0$\nu\beta\beta$ decay experiments 
comes from the fact that - without prevention - at sea level cosmic radiation produces  various
radioactive isotopes of which \cosix\ is most detrimental due to its long half life 
of 5.3 years. Being a contaminant in Ge, \cosix\ is still removable by zone refining, and thus the
clock for its production starts to tick at the end of the last zone refining step. 
Table~\ref{time} shows the time needed by a big supplier for the different processing steps from zone 
refining to the detector end product. 
For optimum
conditions, this implies that detector production, at least, must be done underground. Until such
a facility is available, it might be possible to select suppliers which can store the crystals
underground in the periods where the manufacturing process is interrupted.   

\subsubsection{Loss of \geenr\ material during processing}

The \IGEX\ collaboration has lost about 5\% of enriched \geenr\ material during the various
steps of processing \cite{maj03} and is aiming now at less than  3\%. At present,
we are not able to provide a reliable estimate.

\newpage
 
\section {R\&D program}
\label{sec:RandD}

The schedule for the R\&D projects reflects the different phases of the experiment: 
there will be a relatively short phase of R\&D aiming at optimizing the experimental setup
and the performance of the existing enriched detectors in experiment, followed by a second 
intermediate phase which focuses on the optimum design for the new enriched detectors, on their 
fabrication and a further enhancement of the  performance of the whole setup including the 
instrumentation of  the liquid argon. Nevertheless, much work for the intermediate term R\&D will 
be done in parallel with the short term activity and will start immediately. 

\subsection{Mechanical engineering}

A short term project is the design of the interlock system between the vessel and the outside world. 
This system is closely connected with
the suspension system for up to 50 natural and enriched Ge diodes. An important
requirement is the possibility to exchange a small part of the diodes without removal of
the whole detector array from the vessel.  The suspension system has to solve 
contradicting requirements: while mass reduction is the first goal,
it, nevertheless, must be rigid and  guarantee a well defined orientation and 
location of the diodes.
This task is becoming even more difficult in view of the  
constraint that only materials of established radiopurity  are allowed for construction.

Further short term mechanical R\&D has to address the design of a system which allows to
introduce temporarily radioactive sources into the vessel for periodically establishing the
energy calibration of the Ge diodes. The system must allow one to position
these sources at various precisely known locations so that all detectors are well 'illuminated' 
and the resulting spectra can be used in the comparison with the Monte Carlo simulations of the 
detector array. Obviously,
also this project is closely linked to the interlock and suspension system. Possible solutions are
permanently installed plastic tubes through which  the sources are inserted and removed, or
an internal manipulator system whose remotely controlled arms  move the sources from their shielded
location to the detectors and back.

\subsection{Electronic engineering}

Stable yet ultra-low mass contacts at the crystal are the prerequisite for a reliable low-noise 
operation of the diode. The optimum solution might be ultrasonic bonding with aluminum or gold wires 
of typically 20~$\mu$m diameter. This method is also used  for contacting segmented 
Ge detectors, and first tests with a dummy Ge crystal have yielded encouraging results.
The end of the delicate bond wire would be attached to the signal or high
voltage cable which must, however, be fixed with respect to the crystal in order to avoid
stress onto the delicate bond wire.

The readout cable should be flexible, even at LN temperature, hold high voltages up to 4 or 5~kV, 
exhibit a low impedance for maintaining the pulse characteristics in the time domain, and be well 
shielded to minimize cross talk; in addition it should consist of radiopure materials only.   
First tests showed that Kapton cables with copper traces represent a promising candidate. 
More work has to be done to come up with an optimum solution.

In cryostat-housed Ge detectors, the FET of the preamplifier stage is usually mounted close to
the diode on the cool finger. A little piece of plastic between cool finger and FET serves as thermal 
impedance and makes the  FET run at about 20$^\circ$C above the LN temperature at optimum noise 
performance. This approach is presumably not possible in LN, and it has to be investigated if there
are commercial FETs available which are suited to run at LN temperature, or a
dedicated FET has to be designed which does not exhibit carrier freeze out. Since the cable between FET and
preamplifier might be still several meters long, the most straightforward way to best performance 
will be 
to put not only the FET but the whole preamplifier close to the diode, which - on a few mm$^2$
large silicon die - can be realized by a fully integrated VLSI solution. The \AGATA\ collaboration has 
reported encouraging progress in this field, and R\&D on such a device can be done at the Heidelberg 
ASIC laboratory where the MPI Heidelberg has developed already a VLSI readout ASIC for silicon strip 
detectors. 

\subsection{Monte Carlo simulations}

An urgent topic is the simulation of the planned final vessel setup in order to
arrive at a full understanding of its shielding power, and to optimize the use of
shielding materials. Monte-Carlo simulations will be used also for the layout of the
muon shield including the water Cherenkov detector.

As soon as the vessel layout is fixed, a detailed Monte Carlo description of
the experimental setup has to be developed  including the database which documents
the evolution of the experimental setup with time.

Obviously, the present simulations have focused on the most important backgrounds in the
2~MeV region only. However,
it is desirable to study also  rare processes in order to deepen the quantitative
understanding of the observed background at all energies. This will be of prime importance for dark
matter studies for which the low energy part of the measured spectrum has to be understood
quantitatively.

\subsection{Validation of materials}

Screening of materials to be used for shielding, support and contacts of the Ge diodes 
is of utmost importance. 
So far, these investigations have profited very much from
the most sensitive GeMPI setup at LNGS. 
Present measurements include the determination of the radioactivity of several samples
of commercially available lead which could be used for the vessel shielding.
The assembly of a similar second station is in progress. In addition, with the termination 
of the LENS prototype phase, the "Low background LENS facility" might be used as powerful 
large volume screening device.

Low background measurement devices of a different setup, exhibiting e.g. multi detector
arrays, are installed at the Baksan underground laboratory and use of these
facilities will be possible by the Russian signers of this Letter of Intent.

\subsection{Detector R\&D}

Basic detector R\&D will start with a simulation of the field distribution in various variants 
of p-type Ge diodes which will eventually result in an optimized design of the detectors
as well as in a better understanding of the pulse shapes expected for single- and 
multi-site events. 

R\&D with commercial suppliers of Ge diodes will focus on the efficient manufacturing of
axially segmented p-type detectors. These devices are easier to build than n-type diodes, and
their robust outside dead-layer does both ease the handling and provide an integrated shield against 
$\alpha$ particles which outweighs the loss of sensitive volume.
Tests with radioactive sources will establish how much
information can reliably be extracted from pulse shapes, and how well internal background processes
can be suppressed by the anti-coincidence between different segments. This experimental program will be
accompanied by Monte Carlo simulations. 

As outlined at several occasions above, the fabrication of new enriched Ge diodes should 
be done preferentially underground. This way,  not only cosmogenic contaminations can 
be minimized but complete control of the cleanliness of the production cycle becomes
possible, too. In addition, as known from estimates within the \AGATA\ collaboration, our own production
might be more cost-effective than ordering from a commercial supplier. A realization of this
plan would imply a long term engagement and needs to be prepared with a most detailed R\&D plan.
At present, it is premature to outline such a plan. However, it is in this context interesting
to know that similar considerations exist within the \AGATA\ collaboration, and a  merge of the
interested parties might turn out to be beneficial for both sides with respect to sharing of resources, 
know-how transfer and  training for the fabrication of Ge diodes.

Eventually, an extension of Ge technology to thin and highly segmented Ge diodes might open
a new and perhaps more powerful way for the suppression of both internal and external backgrounds.

\subsection{Germanium enrichment}
\label{sec:gecl4}

Currently the enrichment process is performed with gas centrifuges using GeF$_4$. While this is
a well proven technology there are some disadvantages: the chemical treatment for GeF$_4$ is
relatively expensive and there could be contaminations  from uranium since
the F$_2$ gas used is typically recycled from UF$_6$ enrichments. This contamination will 
accumulate in the separation  step for $^{76}$Ge.

Researchers at the Institute of Molecular Physics of the Kurchatov Institute have developed
a new Ge isotope selection technique based on GeCl$_4$. This method has
been established for the production of a few grams of  $^{76}$Ge. GeCl$_4$ is a commercially
available compound with high chemical purity, and the 
disadvantages mentioned for GeF$_4$ are avoided. The main
limitation of this approach stems from the fact that chlorine has two stable
isotopes ($^{35}$Cl and $^{37}$Cl) and the heavier one has  only an
abundance of 24\%. The maximum possible enrichment fraction amounts to about 40\%.
For higher enrichments the GeF$_4$ technology has to be used which would  require
in this case much less material and would be more efficient.

The  extension of the GeCl$_4$ technology to produce a few
kg of enriched material is one of the foreseen R\&D topics. 
Its isotope selection and economical efficiency will be studied in detail. In addition 
it is planned to develop measurement techniques to identify the U/Th contamination in
the raw and enriched germanium material. Once established this will be applied for
monitoring the separation process.

\subsection{Instrumentation of liquid argon}


Scintillation light in liquid argon is well 
established \cite{kub79,dok90,hit83,cen99}.
Its potential for internal and external background rejection can be 
realized only if the UV light is efficiently collected. 
Cosmic ray muons create large signals and thus can be vetoed readily. 
In order to achieve an efficient anti-coincidence shield for 
cosmogenic activities such as internal $^{60}$Co or $^{68}$Ge, or
external gammas from $^{208}$Tl, an
effective threshold of typically 100~keV should be achieved.
An additional requirement is that materials for shifting and 
guiding the scintillation photons to the photo sensors, 
as well as the photo sensors 
themselves must not augment the background signal in the 
germanium detectors. With instrumented LAr, a densely packed
Ge diode array is not the optimum detector arrangement
since the diodes  are an obstacle for the light collection.
 The optimum geometry has to be resolved with
Monte Carlo studies.  

\begin{figure}[h]
\begin{center}
\includegraphics[width=10cm]{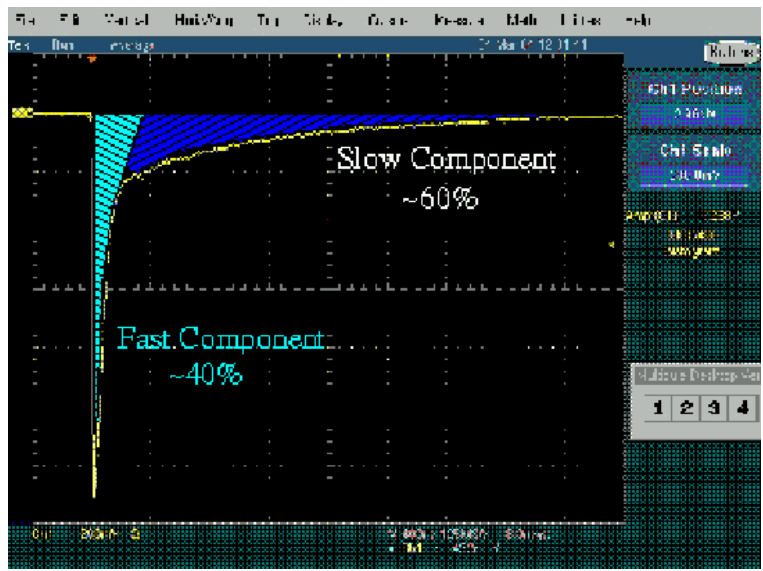}
\end{center}
\caption{\label{larsc}
Measured response to a muon crossing the LAr volume
with the characteristic fast and slow pulse component.
}
\end{figure}

The scintillation yield of liquid argon is 40,000 photons per MeV 
and the wavelength of the emitted photons is 128~nm. 
The strategy  to detect
these photons is to shift them to a wavelength of about
400~nm and to transport them via specular reflection
to photomultiplier tubes immersed in liquid argon.  
A transparent wavelength shifter developed for IceCUBE \cite{res04}
combined with an efficient reflector foil \cite{mot04} appear as 
an attractive solution. First experimental tests have 
started using a PMT (ETL 9367 KFLB) immersed in liquid
argon combined with wavelength shifting and reflector foils.
Fig.~\ref{larsc} shows the recorded pulse shape for muons.
Next steps in the R\&D programme comprise the operation 
of a germanium crystal
in anti-coincidence with the liquid argon scintillation
and measurements of photo-electron yields for different 
wavelength shifting materials.
\clearpage

\newpage

\section{Safety}

The present status of the proposed experiment does not yet allow a detailed safety discussion.
On the other hand, the major generic safety aspects are known, and this chapter serves to discuss
them and to prove that safety aspects are observed already in the early planning stage of the
experiment.

Tables~\ref{isiec} and \ref{isiec1} identify potential hazards 
which may be caused by the installation of the proposed  experiment. 
\begin{table}[ht]
\begin{center}
\caption{\label{isiec} Initial safety information for the proposed experiment, 
excluding the construction phase.
		       }
\vskip-4truemm
\item \begin{tabular}{@{}l}
\hline
\hline
SPOKESPERSON:...... tbd ~~~~~~~~~~~~~~~~~~~~~~~~~~~~~GLIMOS:...... tbd  \\
\hline
\hline
(1) GASES, LIQUIDS, CRYOFLUIDS (used in detectors or kept nearby) \\
\hline
\end{tabular}

\begin{tabular}{lllll}
Device Type & Fluid & Volume & Abs. Press. & Max Flow\\
\hline
~~$>$ Exp. tank & LN or LAr  & $<$400~m$^3$ & 1.2 atm &  \\
~~$>$ Exp. tank & water      & $<$500~m$^3$ & 1.0 atm &  \\
\end{tabular}
\vskip-2truemm

\item \begin{tabular}{@{}l}
\hline
\hline
(2) OTHER CHEMICALS \\
~~~~(Toxic/Corrosive/Flammable solvents, additives etc): \\
\hline
~~$>$ depending on type, vessel isolation may emit toxic gases if burning\\
~~$>$ small amounts of cleaning fluids like alcohol \\

\hline
\hline
(3) ELECTRICITY \hskip9.5truecm \phantom{a} \\
\hline
~~MAGNETS: $>$ NONE \\
\hline
~~High Voltage ($>$1~kV) \\
\end{tabular}

\begin{tabular}{llllc}
\hline
Detector Type & Voltage & Current & Stored Energy & No of HV channels \\
\hline
$>$ Ge diode      & $<$5~kV & $<$1~nA &               &   $<$50   \\
$>$ PM            & $<$3~kV & $<$3~mA &               & {\ord O}(100)   \\
\hline
\end{tabular}

\vskip-4truecm
\item \begin{tabular}{@{}l}
~~SHORT-CIRCUIT current $>$5~mA for $>$50V possible anywhere? $>$ (NO) \\
\hline

~~POWER dissipated by all electronics \\
~~a) on detectors: $>$ negligible~~~b) off detectors: $>$ to be specified\\
\hline
~~SPECIAL GROUNDING REQUIREMENTS?~~~~$>$ NONE\\
\hline
\hline
\end{tabular}
\end{center}
\end{table}
The still incomplete tables follow the initial safety information requested for experiments
at CERN \cite{cedms}.  Information on the spokesperson  and the group leader in matter of safety (GLIMOS) 
is not yet available.
Table \ref{isiec} shows that an impact on the environment by toxic or corrosive materials 
is excluded. 
No flammable gases and liquids are used. However,
certain types of insulation under consideration (e.g. polyurethane) do emit toxic gases if they are 
burning. Replacements are under investigation. 
The radioactive sources needed for the energy calibration of the Ge detectors will be all 
encapsulated, and their activity will be low enough so that no regulations beyond the standard radiation 
protection rules are needed.  
\begin{table}[ht]
\begin{center}
\caption{\label{isiec1} Initial safety information for the proposed experiment, 
excluding the construction phase (continued).
		       }
\vskip-4truemm
\item \begin{tabular}{@{}l}
\hline
\hline
(4) LIFTING AND HANDLING \hskip7.3truecm      \phantom{a}\\
\hline
~~Weight of heaviest single piece to install?~~~~$>$ to be specified\\
~~Specially designed handling equipment?~~~~~~ For which max. weight?\\
\hline
\hline
(5) VACUUM TANK, PRESSURE TANK, CRYO-TANK  \\
\end{tabular}

\begin{tabular}{llll}
\hline
Tank & Abs. pressure & Volume & Weakest part of wall \\
\hline
$>$ Cryo-tank (exp.)   & 1.2 atm.&   $<$250~m$^3$        &     \\
$>$ Cryo-tank1 (storage) & to be specified    &                       &     \\
$>$ Cryo-tank2 (storage) & to be specified    &                       &     \\
\end{tabular}
\item \begin{tabular}{@{}l}
\hline
\hline
(6) IONIZING RADIATION (radioact. sources, depleted uranium, etc.)~~~~\phantom{a} \\
\hline
~~Radioactive source~~~~~~~~~$>$ for calibration, to be specified \\
\hline
\hline

(7) NON-IONIZING RADIATION (Laser, UV light, microwaves, rf) \\
\hline
~~$>$ UV laser for PMT calibration \\
~~$>$ possibly UV light for sterilizing water \\
\hline
\hline

(8) OTHER HAZARDS  \phantom{a} \\
\hline
~~$>$ suffocation \\
~~$>$ electric shorts induced by water leak \\
\hline
\hline
(9) RISK ANALYSIS~~~~~~~~~~~~~\phantom{a} \\
\hline
~~$>$ to be done for big LN/LAr/water vessel \\
\hline
\hline
\end{tabular}
\end{center}
\end{table}

The major potential hazard is due to the big cryostat of the experiment which will
contain a maximum of 250~m$^3$ of liquid nitrogen (LN) or liquid argon (LAr). Table~\ref{lgp} gives
a compilation of the physical properties of these cryogenic liquids \cite{tis09}.
\begin{table}[ht]
\begin{center}
\caption{\label{lgp} Physical properties of liquid nitrogen and argon}
\begin{tabular}{@{}lrrl}
\\
\hline
\hline
Characteristics & Argon  & Nitrogen & Unit\\
\hline
\hline
Boiling point at 1 bar & -185.5 & -195.8  & $^\circ$C \\
                       & 87.3   &   77.3  & K         \\   
\hline
Density of liquid at boiling point  & 1400 & 810 &  kg/m$^3$ \\
\hline
Liters of gas at 20$^\circ$C produced  & 841 & 693 & liters \\
by 1 liter of liquid, 1 bar \\
\hline
Density at 20$^\circ$C compared to     & 1.4  & 1.0 \\ 
density of air \\
\hline
Latent heat of evaporation for        & 220 & 160 & kJ \\
1 liter of liquid \\
\hline
Ratio of enthalpy of vapor at 20$^\circ$C & 0.7 & 1.14 \\ 
and latent heat of evaporation \\
\hline
\hline
\end{tabular}
\end{center}
\end{table}

The amount of stored LN or LAr corresponds to a gaseous volume of about 
210000 (180000) \cum ; this is about 9 (7) times the volume of Hall A.
The specified daily evaporation is less than 0.5\% of the liquid, or less than 10\% 
 of the volume of Hall A.
At standard operation, the present ventilation system guarantees that 40\% of the
air is exchanged per hour in each of the three halls, or an \lq air-washing\rq\ of the 
total volume of each hall within 2.5 hours.

Argon and nitrogen are inert, non-toxic and non-flammable gases.
However, if they replace part of the oxygen in the atmosphere, severe damage to
human beings can result. This hazard is increased in a confined space like an
underground hall with limited ventilation. 
Thus, as a result of leakage, the hazards will be essentially that of asphyxiation 
and injuries due to the low 
temperature - frostbites, cryo-burns, hypothermia and others.
Similarly, due to low temperature further damage could result to 
structures and equipment impinged on by the cryofluid. 
A risk analysis has to show which measures are needed to prevent a major accident,
and to minimize the risks and effects of a major accident. On the other hand, it
is known \cite{ful88} that the use of a doubly-walled cryogenic vessel with passive
insulation makes it highly unlikely that a major leakage of liquid fluid can 
develop. A study of a super-insulated vessel version has shown that the requirement
to withstand a standard earthquake can be met rather easily \cite{bab03}. 

All other potential hazards are not aggravated due to the fact that the experiment 
will be operated underground. 
It is intended to plan and to build the experiment and its infrastructure following 
the \lq Safety Guide of the \LNGS \rq\ and the safety rules of \CERN .
In particular, it is intended to provide in  due time documentation of the safety relevant 
components. \\
This documentation will include \\
\phantom{a}$\quad$$\bullet$  tank cooling down and filling procedure, \\
\phantom{a}$\quad$$\bullet$  tank emptying procedure, \\
\phantom{a}$\quad$$\bullet$  description/location of relieve and safety valves,\\
\phantom{a}$\quad$$\bullet$  description/location of temperature and pressure sensors,\\
\phantom{a}$\quad$$\bullet$  description/location of gas collector diffusers,\\
\phantom{a}$\quad$$\bullet$  description/location of oxygen monitors,\\
\phantom{a}$\quad$$\bullet$  description of gas control system and its operation,\\
\phantom{a}$\quad$$\bullet$  inventory of installed electrical cables,\\
\phantom{a}$\quad$$\bullet$  description of fire extinguishing system \\
\phantom{a}$\quad$$\bullet$  risk analysis/matrix with respect to LN/LAr/water leakage,\\
\phantom{a}$\quad$$\bullet$  risk analysis/matrix considering earthquake, statics calculations   \\
\phantom{a}$\quad$$\bullet$  risk analysis/matrix considering power black-out,         \\
\phantom{a}$\quad$$\bullet$  risk analysis/matrix considering possibility of burning insulation.         

The overall safety of the experiment will profit from a fully computerized control and monitoring
system with remote access that will record all safety relevant parameters.
The critical parts of this highly reliable and redundant detector safety system will be based on an 
autonomously running programmable logic controller (PLC) front-end.
In case of any alarm condition, the data will be communicated to the \LNGS\ general safety monitoring
system which will trigger the necessary action and contact  the experiment's  on-call duty.
In case of electrical power outage, a UPS module makes it possible to continue data taking or 
to automatically bring  all critical system components into a safe state.

\newpage
%
%

\section{Time schedule}
\label{sec:timeschedule}

The experiment is divided into two phases as shown in Table~\ref{phases}. A third phase is considered.
Phase I
\begin{table}[ht]
\begin{center}
\caption{\label{phases} Phases of the planned experiment}
\vskip2truemm
\begin{tabular}{|l |@{}c@{} | @{}l@{} | @{}l@{} | @{}c@{} | @{}c@{} | @{}c@{} |}
\hline
 & 2005 & 2006 & 2007 & 2008 & 2009 & 2010\\ 
\hline
{\bf Phase I} &&&&&& \\
Construction  &\balf&&&&& \\
Measurement with existing \geenr\ diodes &&\balf&\rule{5mm}{2mm}\hfill \rule{2mm}{2mm}&
\hfill \rule{2mm}{2mm}\hfill \rule{2mm}{2mm}\hfill &
\hfill \rule{2mm}{2mm}\hfill \rule{2mm}{2mm}\hfill &
\hfill \rule{2mm}{2mm}\hfill \rule{2mm}{2mm}\hfill  \\
\hline
{\bf Phase II} &&&&&& \\
Procurement of enriched \geenr\ material &\balf &{\rule{5mm}{2mm}}&  &&& \\
Production of new \geenr\ diodes &&\balf & \balf &&& \\
Measurement with all \geenr\ diodes &&&\balf&\balf&\balf&\balf \\
\hline
\end{tabular}
\end{center}
\end{table}
starts with the construction of the experimental facility, and measurements begin
one year later using both natural Ge diodes as well as the major part of the existing almost 20~kg 
enriched \geenr\
diodes. The actual beginning of Phase II is determined by the availability of funds for the procurement
of enriched \geenr\ material and the fabrication of new probably segmented enriched \geenr\ detectors.
We assume this is the case soon after the construction is finished. 
With the new detectors the running will continue until about hundred (kg$\cdot$years) have been 
accumulated and backgrounds start to show up. The outcome of this Phase II will determine how
Phase III has to be defined.

Table~\ref{tab2} shows in more detail the schedule for the construction period in Phase I. It is
assumed that the funds are secured by the end of the third quarter in 2004. The procurement time
for the vessel accounts for the need to have an European-wide call for tenders. The first
steps towards the procurement of enriched \geenr\ material will be made as soon
as funds are available in order to avoid any loss in time.

\begin{table}
\begin{center}
\caption{\label{tab2} Proposed technical time schedule for Phase I}
\vskip2truemm
\begin{tabular}{|l |@{}c@{} @{}c@{} @{}c@{} @{}c@{} | @{}c@{}  @{}c@{}  @{}c@{}  @{}l@{}|}
\hline
 & 2004 &&&&            2005 & & &  \\ \hline
 & Q1 & Q2 & Q3 & Q4 &  Q1 & Q2 & Q3 & Q4 \\ \hline
{\bf LoI} & \balf &&&&&&& \\ \hline
{\bf Proposal} &&& \balf &&&&&\\ \hline
{\bf Tank} &&&& &&&& \\
Review, optimize & & \balf & \balf & & & & &   \\
Procure     & & & & \balf & & & &   \\
Construct   & & & &      & \balf & \balf & &  \\
Install     & & & &      &       &       & \balf & \\
Cryogenic commissioning &&&& & & & & {\rule{5mm}{2mm}} \\
\hline
{\bf Lock, suspension, calibration} &&&& &&&& \\
Design &&\balf&\balf&\balf&&&&\\ 
Fabricate, test      &&&& &\balf&\balf&& \\
Install                &&&& &&&\balf& \\
\hline
{\bf Muon veto system } &&&& &&&& \\
Design/build            &&&&\balf &\balf&\balf&& \\ 
Install                 &&&& &&&& \balf \\ 
\hline
{\bf Electronics}      &&&& &&&& \\
Frontend R\&D          &&\balf&\balf&\balf &&&&\\ 
Optimize contacts, cables  &&\balf&\balf&\balf &&&& \\
Design/build frontend &&&&\balf &\balf&\balf&& \\ 
Design DAQ/database    &&&&\balf &\balf&&& \\
Install/test DAQ       &&&& &&\balf&& \\
\hline
{\bf Ge diodes} &&&& &&&& \\
Modify existing diodes &&&&\balf &\balf&\balf&&\\ 
Test modified diodes   &&&&      &\balf&\balf&\balf& \\
Simulate segmented diodes &&&\balf&\balf &&&& \\ 
Procure segmented diode &&&& &\balf&&& \\ 
Test diodes in LN and LAr  &\balf&\balf&\balf&\balf &\balf&\balf&\balf& \\
Install diodes             &&&& &&&\multicolumn{2}{r@{}|}{\rule{5mm}{2mm}}  \\
\hline
{\bf Enriched $^{76}$Ge detectors }  &&&& &&&& \\
Procure test batch material &&&&\balf &&&& \\
Validate material           &&&& &\balf&&& \\
Build/test small test diode &&&& &\balf&\balf&& \\
Order more enriched material &&&& &&&\balf&\balf \\
\hline
\end{tabular}
\end{center}
\end{table}

\clearpage

\newpage
\addcontentsline{toc}{section}{\protect\numberline{References}}

\end{document}